%% file: main-paper.tex
\RequirePackage{fix-cm}
\documentclass[twocolumn]{svjour3}          %
\smartqed  %
\usepackage{graphicx}

\usepackage[numbers, compress]{natbib}
\usepackage{eucal}
\usepackage{footnote}
\usepackage{hyperref}
\usepackage{booktabs}
\usepackage[table]{xcolor} 
\usepackage{float}
  
\usepackage{mathptmx}      %
\usepackage{balance}
\usepackage{amssymb}
\usepackage{amsmath}
\usepackage{amsfonts}
\DeclareMathAlphabet{\mathcal}{OMS}{cmsy}{m}{n}
\usepackage{colortbl}
\usepackage{xspace,balance,tabularx,multirow}
\usepackage{flushend}
\usepackage{tikz}
\usepackage{pgfplots}
\usepackage{pgfplotstable}
\usepackage{caption}
\usepackage{subfig}
\usepackage{enumitem}
\usepackage{tablefootnote}
\usepackage{graphicx}
\DeclareUnicodeCharacter{2212}{−}
\usepgfplotslibrary{groupplots,dateplot}
\usetikzlibrary{patterns,shapes.arrows}
\pgfplotsset{compat=newest}
\usepackage[ruled, vlined, linesnumbered]{algorithm2e}

\usepackage[utf8]{inputenc} %
\usepackage[T1]{fontenc}    %
\usepackage{microtype}      %
\usepackage{footnote}

\newcommand{\CS}{\mathcal{C}\xspace}
\newcommand{\AM}{\mathbf{A}\xspace}
\newcommand{\DM}{\mathbf{D}\xspace}

\newcommand{\HM}{\mathbf{H}\xspace}
\newcommand{\IM}{\mathbf{I}\xspace}
\newcommand{\SM}{\mathbf{S}\xspace}
\newcommand{\MM}{\mathbf{M}\xspace}

\newcommand{\PM}{\mathbf{P}\xspace}
\newcommand{\XM}{\mathbf{X}\xspace}
\newcommand{\YM}{\mathbf{Y}\xspace}
\newcommand{\ZM}{\mathbf{Z}\xspace}
\newcommand{\BM}{\mathbf{B}\xspace}

\newcommand{\QM}{\mathbf{Q}\xspace}
\newcommand{\RM}{\mathbf{R}\xspace}
\newcommand{\WM}{\mathbf{W}\xspace}
\newcommand{\UM}{\mathbf{U}\xspace}
\newcommand{\VM}{\mathbf{V}\xspace}
\newcommand{\TM}{\mathbf{T}\xspace}
\newcommand{\FM}{\mathbf{F}\xspace}

\newcommand{\OmegaM}{\boldsymbol{\Omega}\xspace}

\newcommand{\PN}{\mathbf{P}_{N}\xspace}
\newcommand{\stitle}[1]{\vspace{1mm}\noindent{\bf #1.}}
\newcommand{\transpose}{^\mathsf{T}}
\DeclareMathOperator*{\argmin}{argmin}

\newcommand{\ie}{{\it i.e.},\xspace}

\newcommand{\extendgpu}{\texttt{ANCKA-GPU}\xspace}
\newcommand{\extendagc}{\texttt{ANCKA}\xspace}

\newcommand{\mainalgo}{\texttt{AHCKA}\xspace}
\newcommand{\mhc}{\texttt{CalMHC}\xspace}
\newcommand{\bcm}{\texttt{InitBCM}\xspace}
\newcommand{\kmeans}{\texttt{k-means}\xspace}
\newcommand{\discr}{\texttt{Discretize}\xspace}
\newcommand{\discrgpu}{\texttt{Discretize-GPU}\xspace}

\newcommand{\hncut}{\texttt{HNCut}\xspace}
\newcommand{\hyperadj}{\texttt{HyperAdj}\xspace}
\newcommand{\kahypar}{\texttt{KaHyPar}\xspace}

\newcommand{\gnmfl}{\texttt{GNMF}\xspace} %
\newcommand{\jnmf}{\texttt{JNMF}\xspace}
\newcommand{\grac}{\texttt{GRAC}\xspace}
\newcommand{\arw}{\texttt{ACMin}\xspace}
\newcommand{\arwc}{\texttt{ACMin-C}\xspace}
\newcommand{\arws}{\texttt{ACMin-S}\xspace}
\newcommand{\athncut}{\texttt{ATHNCut}\xspace}
\newcommand{\athyperadj}{\texttt{ATHyperAdj}\xspace}
\newcommand{\atkahypar}{\texttt{ATKaHyPar}\xspace}
\newcommand{\atmetis}{\texttt{ATMetis}\xspace}

\newcommand{\metis}{\texttt{Metis}\xspace}
\newcommand{\atncut}{\texttt{ATNCut}\xspace}
\newcommand{\ncut}{\texttt{NCut}\xspace}
\newcommand{\gnmf}{\texttt{GNMF}\xspace}
\newcommand{\agc}{\texttt{AGCGCN}\xspace}
\newcommand{\fgc}{\texttt{FGC}\xspace}
\newcommand{\grace}{\texttt{GRACE}\xspace}
\newcommand{\gracegpu}{\texttt{GRACE-GPU}\xspace}
\newcommand{\cugraph}{\texttt{SMM-GPU}\xspace}

\newcommand{\omac}{\texttt{O2MAC}\xspace}
\newcommand{\hdmi}{\texttt{HDMI}\xspace}
\newcommand{\magc}{\texttt{MAGC}\xspace}
\newcommand{\mcgc}{\texttt{MCGC}\xspace}

\newcommand{\cesna}{\texttt{CESNA}\xspace}
\newcommand{\infomap}{\texttt{Infomap}\xspace}
\newcommand{\louvain}{\texttt{Louvain}\xspace}
\newcommand{\aggcluster}{\texttt{HAC}\xspace}

\newcommand{\kmqi}{\texttt{k-MQI}\xspace}
\newcommand{\knibble}{\texttt{k-Nibble}\xspace}

\newcommand{\revision}[1]{{{#1}}}
\newcommand{\revcolor}{black}
\newcommand{\extension}[1]{{{#1}}}

\newcommand{\eat}[1]{{}}
\newcommand{\inReport}[1]{{}}
\newcommand{\inPaper}[1]{{#1}}

\makeatletter

\newcommand*{\@rowstyle}{}

\newcommand*{\rowstyle}[1]{%
  \gdef\@rowstyle{#1}%
  \@rowstyle\ignorespaces%
}

\newcolumntype{=}{%
  >{\gdef\@rowstyle{}}%
}

\newcolumntype{+}{%
  >{\@rowstyle}%
}

\makeatother

\newenvironment{customlegend}[1][]{%
    \begingroup
    \csname pgfplots@init@cleared@structures\endcsname
    \pgfplotsset{#1}%
}{%
    \csname pgfplots@createlegend\endcsname
    \endgroup
}%
\def\addlegendimage{\csname pgfplots@addlegendimage\endcsname}

\begin{document}
\include{data}

\title{A Versatile Framework for Attributed Network Clustering via K-Nearest Neighbor Augmentation
}

\author{Yiran Li \and Gongyao Guo  \and Jieming Shi \and Renchi Yang \and Shiqi Shen \and Qing Li \and Jun Luo %
}

\institute{Yiran Li, The Hong Kong Polytechnic University \at
              \email{yi-ran.li@connect.polyu.hk}           %
        \and
           Gongyao Guo, The Hong Kong Polytechnic University \at
              \email{gongyao.guo@connect.polyu.hk}
              \and
           Jieming Shi, The Hong Kong Polytechnic University  \at
              \email{jieming.shi@polyu.edu.hk }(Corresponding author)
        \and
           Renchi Yang, Hong Kong Baptist University \at
              \email{renchi@hkbu.edu.hk}
        \and Shiqi Shen, WeChat Tencent \at 
        \email{shiqishen@tencent.com}
        \and 
        Qing Li, The Hong Kong Polytechnic University \at \email{csqli@comp.polyu.edu.hk}
        \and 
        Jun Luo, Logistics and Supply Chain MultiTech R\&D Centre \at \email{jluo@lscm.hk}
}

\date{Received: date / Accepted: date}

\maketitle
\begin{abstract}
Attributed networks containing entity-specific information in node attributes are ubiquitous in modeling social networks, e-commerce, bioinformatics, etc. 
Their inherent network topology ranges from simple graphs to hypergraphs with high-order interactions and multiplex graphs with separate layers.
An important graph mining task is node clustering, aiming to partition the nodes of an attributed network into $k$ disjoint clusters such that intra-cluster nodes are closely connected and share similar attributes, while inter-cluster nodes are far apart and dissimilar. 
It is highly challenging to capture multi-hop connections via nodes or attributes for effective clustering on multiple types of attributed networks.

In this paper, we first present \mainalgo as an efficient approach to {\em attributed hypergraph clustering} (AHC).
\mainalgo includes a carefully-crafted $K$-nearest neighbor augmentation strategy for the optimized exploitation of attribute information on hypergraphs, a joint hypergraph random walk model to devise an effective AHC objective, and an efficient solver with speedup techniques for the objective optimization.
The proposed techniques are extensible to various types of attributed networks, and thus, we develop \extendagc as a versatile attributed network clustering framework, capable of {\em attributed graph clustering} (AGC),  {\em attributed multiplex graph clustering} (AMGC), and AHC. 
Moreover, we devise \extendgpu with algorithmic designs tailored for GPU acceleration to boost efficiency. 
We have conducted extensive experiments to compare our methods with \revision{19} competitors
on 8 attributed hypergraphs, \revision{16} competitors on 6 attributed graphs, and \revision{16} competitors on 3 attributed multiplex graphs, all demonstrating the superb clustering quality and efficiency of our methods.

\end{abstract}

\input{introduction}

\input{preliminaries}

\input{methodology}

\input{extension}
\input{experiment}

\input{related_work}

\input{conclusion}

\begin{acknowledgements}
The work described in this paper was fully supported by grants from the Research Grants Council of the Hong Kong Special Administrative Region, China (No. PolyU 25201221, PolyU 15205224, PolyU 15200023).
Jieming Shi is supported by NSFC No. 62202404.
Renchi Yang is supported by the NSFC YSF grant (No. 62302414) and Hong Kong RGC ECS grant (No. 22202623).
Jun Luo is supported by The Innovation and Technology Fund (Ref. ITP/067/23LP).
This work is supported by
Tencent Technology Co., Ltd. P0048511.

\end{acknowledgements}

\bibliographystyle{spbasic}      %
\bibliography{sample}  %

\end{document}

%% file: introduction.tex
\section{Introduction}\label{intro}
\extension{An attributed network contains a network topology with attributes associated with nodes. 
Representative types of attributed networks include attributed graphs, attributed hypergraphs, and attributed multiplex graphs. 
Given an attributed network $\mathcal{N}$, node clustering is an important task in graph mining, which aims to divide the $n$ nodes of $\mathcal{N}$ into $k$ disjoint clusters, such that nodes within the same cluster are close to each other in the network topology and similar to each other in terms of attribute values. 
Clustering on  attributed networks finds important applications in 
biological  analysis \cite{duHybridClusteringBased2019}, online  marketing \cite{xuModelbasedApproachAttributed2012}, social network~\cite{yang2013community,ShiMWC14}, Web analysis~\cite{whangMEGAMultiviewSemisupervised2020}, etc.}

\extension{In this work, we present \extendagc, an effective and efficient attributed network clustering method that is versatile to support attributed hypergraph clustering (AHC), attributed graph clustering (AGC), and attributed multiplex graph clustering (AMGC). \extendagc subsumes our previous work \mainalgo \cite{Li2023EfficientAE} that is dedicated to AHC. In what follows, we first elaborate on AHC and then generalize to AGC and AMGC.} 

In a hypergraph, each edge can join an arbitrary number of nodes, referred to as a {\em hyperedge}. The hyperedge allows a precise description of multilateral relationships between nodes, such as collaboration relationships of multiple authors of a paper, interactions among proteins~\cite{GaudeletMP18}, products purchased together in one shopping cart,  transactions involving multiple accounts \cite{wu2021towards}. In practice, nodes in hypergraphs are often associated with many attributes, e.g., the academic profile of authors and the descriptive data of products. The AHC problem is to divide the $n$ nodes in such an attributed hypergraph into $k$ disjoint clusters such that nodes within the same cluster are close to each other with high connectedness and homogeneous attribute characteristics. 
{AHC finds numerous real-life applications in community discovery~\cite{huangHigherOrderConnection2021}, organization structure detection~\cite{duHybridClusteringBased2019}, Web query analysis~\cite{whangMEGAMultiviewSemisupervised2020}, 
biological analysis~\cite{wu2020cancersample}, etc. As another example, AHC can cluster together academic publications with high relevance by considering co-authorship hyperedges and keyword attributes in academic hypergraphs~\cite{ fanseukamhouaHyperGraphConvolutionBased2021}}.

Effective AHC computation is a highly challenging task, especially for large attributed hypergraphs with millions of nodes. 
First, nodes, hyperedge connections, and attributes are heterogeneous objects with inherently different traits, whose information cannot be seamlessly integrated in a simple and straightforward way. 
Second, as observed in previous works on simple graphs \cite{zhouClusteringLargeAttributed2010,yangEffectiveScalableClustering2021}, higher-order relationships between nodes and node-attribute associations are crucial for clustering. 
However, computing such multi-hop relationships and associations via hyperedges usually with more than two nodes in attributed hypergraphs is rather difficult due to the complex hypergraph structures and prohibitive computational overheads (up to $O(n^2)$ in the worst case). 

In the literature, a plethora of clustering solutions \cite{ kahypar10.1145/3529090,hayashi_hypergraph_2020,kumarHypergraphClusteringModularity2018} are developed for plain hypergraphs. These methods overlook attribute information, leading to severely compromised AHC result quality. Besides, a large body of research on attributed graph clustering is conducted, resulting in a cornucopia of efficacious techniques \cite{xuModelbasedApproachAttributed2012,yangEffectiveScalableClustering2021}. However, most of these works cannot be directly applied to handle large attributed hypergraphs with more complex and unique structures. Inspired by the technical advances in the above fields, a number of efforts have been made towards AHC computation in the past years. 
{The majority of AHC methods rely on non-negative matrix factorization \cite{ duHybridClusteringBased2019, whangMEGAMultiviewSemisupervised2020}}, which requires numerous iterations of expensive matrix operations and even colossal space costs of materializing $n\times n$ dense matrices. Particularly, none of them take into account the higher-order relationships between nodes, thereby limiting their result utility. {The state-of-the-art approach \grac \cite{fanseukamhouaHyperGraphConvolutionBased2021} extends {\em graph convolution} \cite{welling2016semi} to hypergraphs}, indirectly incorporating higher-order relationships of nodes and attributes for clustering.
Notwithstanding its enhanced clustering quality, \grac runs in $O(n^2)$ time as an aftermath from costly graph convolution and SVD operations, which is prohibitive for large hypergraphs.
To recapitulate, existing AHC approaches either yield sub-optimal clustering results or incur tremendous computational costs, rendering them impractical to cope with large attributed hypergraphs with millions of nodes.

Given the above, can we combine and orchestrate hypergraph topology and attribute information in an optimized way for improved clustering quality while achieving high scalability over large attributed hypergraphs? We offer a positive answer by presenting \mainalgo (\underline{A}ttributed \underline{H}ypergraph \underline{C}lustering via \underline{K}-nearest neighbor \underline{A}ugmentation), a novel AHC approach that significantly advances the state of the art in AHC computation. 
\mainalgo surpasses  existing solutions through several key techniques. The first one is a $K$-nearest neighbor (KNN) augmentation scheme, which augments the original hypergraph structure with a KNN graph containing additional connections constructed by adjacent nodes with $K$ highest attribute similarities. This is inspired by a case study on a real dataset manifesting that incorporating all-pairwise node connections via attributes or none of them jeopardizes the empirical clustering quality. Second, \mainalgo formulates the AHC task as a novel optimization problem based on a joint random walk model that allows for the seamless combination of high-order relationships from both the hypergraph and KNN graph. Further, \mainalgo converts the original NP-hard problem into an approximate matrix trace optimization and harnesses efficient matrix operations to iteratively and greedily search for high-quality solutions.
Lastly, \mainalgo includes an effective initialization method that considerably facilitates the convergence of the optimization process using merely a handful of iterations. 
{We conduct extensive experiments on  attributed hypergraph data in different domains. Compared with baselines,  \mainalgo exhibits superior performance in both clustering quality and efficiency.} For instance, on the Amazon dataset with 2.27 million nodes, \mainalgo gains over $10$-fold speedup and a significant improvement of $4.8\%$ in clustering accuracy compared to state-of-the-art. Our work  \mainalgo has been published in~\cite{Li2023EfficientAE}.

\extension{In addition to attributed hypergraphs, attributed graphs and attributed multiplex graphs are prevalent in real-world scenarios, such as social networks \cite{Peng2023UnsupervisedMG} and citation networks \cite{Pensky2021ClusteringOD}. 
Different from hypergraphs that allow more than two nodes to form an edge, in a graph, an edge connects exactly two nodes. 
A multiplex graph consists of multiple layers of graphs with a shared set of nodes, and different graph layers represent node connections from different perspectives or domains, e.g., different types of relationships or relations formed in different time frames or spaces \cite{Peng2023UnsupervisedMG, Pensky2021ClusteringOD}.
Attributed graph clustering (AGC) is one of the most significant graph mining problems, extensively studied in the literature \cite{xuModelbasedApproachAttributed2012, yangEffectiveScalableClustering2021}, with many applications, e.g., community detection in social networks \cite{Fortunato2009CommunityDI} and functional cartography of metabolic networks \cite{Guimer2005FunctionalCO}.
Furthermore, a rich collection of studies on attributed multiplex graph clustering (AMGC) also exists in \cite{Pan2021MultiviewCG, Lin2021MultiViewAG, Jing2021HDMIHD, Fan2020One2MultiGA}, to support important applications, e.g., biological analysis \cite{Pensky2021ClusteringOD}, community detection \cite{Peng2023UnsupervisedMG} and social analysis \cite{DeFord2017SpectralCM}.
A previous general framework~\cite{FanseuKamhoua2022GRACEAG} relies on expensive graph convolutions to support various clustering tasks.}

\extension{In this work, we extend \mainalgo for AHC to a versatile framework \extendagc that can efficiently handle attributed \underline{N}etwork clustering tasks (AHC, AGC, and AMGC) to produce high-quality clusters on large data. \extendagc inherits the powerful KNN augmentation scheme and the formulation of clustering objective in \mainalgo. We further develop a generalized joint random walk model in \extendagc with proper transition matrices to support random walks on KNN augmented hypergraphs, graphs, and multiplex graphs simultaneously.
Efficient optimization techniques are applied in \extendagc to retain the advantage of high efficiency for clustering. 
Despite the superior efficiency, clustering million-scale datasets with \extendagc can still take dozens of minutes. 
Moreover, 
after observing the limited speedup ratio by increasing the number of CPU threads used,  we pinpoint the efficiency bottlenecks and design the GPU-accelerated \extendgpu,
to boost the efficiency to another level, especially on large-scale datasets. \extendgpu consists of GPU-based optimization techniques and KNN construction procedures to speed up.
We have conducted extensive experiments to compare \extendagc with {16} competitors on various attributed graphs and {16} competitors on attributed multiplex graphs. 
In all three tasks, \extendagc obtains superior performance regarding both clustering quality and efficiency. The GPU implementation \extendgpu further reduces time costs significantly, often by an order of magnitude on large datasets.}

We summarize the contributions of this work as follows:
\vspace{-\topsep}
\begin{itemize}[leftmargin=*]
\item We devise a KNN augmentation scheme that exploits attributes to augment the original hypergraph structure in a cost-effective manner.
\item We formulate the AHC task as an optimization with the objective of optimizing a quality measure based on a joint random walk model over the KNN augmented hypergraph.
\item We propose a number of techniques for efficient optimization of the objective, including a theoretically-grounded problem transformation, a greedy iterative framework, and an effective initialization approach that drastically reduces the number of iterations till convergence.
\item \revision{We justify  the application of KNN augmentation to various types of networks, generalize the techniques, and design a versatile method \extendagc to efficiently perform AHC, AGC, and AMGC and produce high-quality clusters.}
\item \revision{We develop \extendgpu with customized GPU kernels to improve the efficiency further with a series of GPU-based optimizations while maintaining clustering quality.}
\item \revision{The excellent performance of \extendagc is validated by comprehensive experiments against 19 AHC competitors, 16 AGC competitors, and 16 AMGC competitors, over real-world datasets.}

\end{itemize}

The remainder of this paper is structured as follows: Section \ref{sec:pre} introduces the preliminaries of AHC, AGC, and AMGC. Section \ref{sec:AuHC} outlines the KNN augmentation strategy and random walk scheme for AHC, along with the AHC clustering objective. Section \ref{sec:objective} offers a theoretical analysis of the proposed AHC method \mainalgo, while Section \ref{sec:algo} details the algorithmic procedures of \mainalgo. Section \ref{sec:extendagc} presents the versatile \extendagc framework for AHC, AGC, and AMGC. Section \ref{sec:extendgppu} discusses GPU-based techniques for enhancing clustering efficiency in \extendgpu. Section \ref{sec:experiment} provides a comprehensive experimental evaluation. Section \ref{sec:relatedwork} reviews relevant literature, and Section \ref{sec:conclusion} concludes the paper.

%% file: preliminaries.tex
\section{Preliminaries}\label{sec:pre}

\revision{
\stitle{Attributed Network}
Let $\mathcal{N}=(\mathcal{V}, \mathcal{E}, \XM)$ be an attributed network, where $\mathcal{V}$ is the node set with cardinality $|\mathcal{V}|=n$, $\mathcal{E}$ is the edge (or hyperedge) set with cardinality $|\mathcal{E}|=m$, and $\XM \in \mathbb{R}^{n\times d}$ represents a node attribute matrix. A node $v_j\in \mathcal{V}$ has degree  $\delta(v_j)$, which is the number of edges (or hyperedges) incident to $v_j$. Each node $v_j$ in $\mathcal{V}$ is associated with a $d$-dimensional attribute vector, denoted as $\XM[j]$, i.e., the $j$-th row of the node attribute matrix $\XM$. We consider three types of attributed networks $\mathcal{N}$, including attributed hypergraphs $\mathcal{H}$, attributed graphs $\mathcal{G}$, and attributed multiplex graphs $\mathcal{G}_M$, characterized by different nature of $\mathcal{E}$.}

\revision{
\vspace{0.5pt}\noindent \textbf{Attributed Hypergraph} is denoted by $\mathcal{H}=(\mathcal{V}, \mathcal{E}, \XM)$. $\mathcal{E}$ is the set of $m$ hyperedges where each $e_i\in \mathcal{E}$ is a subset of $\mathcal{V}$ containing at least two nodes.} A hyperedge $e_i$ is said to be incident with a node $v_j$ if $v_j\in e_i$. We denote by $\HM\in \mathbb{R}^{m\times n}$ the incidence matrix of hypergraph $\mathcal{H}$, where each entry $\HM[i,j]=1$ if $v_j\in e_i$, otherwise $\HM[i,j]=0$. Let diagonal matrices $\DM_V\in \mathbb{R}^{n\times n}$ and $\DM_E\in \mathbb{R}^{m\times m}$ represent the degree matrix and hyperedge-size matrix of $\mathcal{H}$, where the diagonal entry $\DM_V[j,j]=\delta(v_j)$ for $v_j\in \mathcal{V}$ and $\DM_E[i,i]=|e_i|$ for $e_i\in \mathcal{E}$, respectively. 
Figure \ref{fig:ahg} shows an attributed hypergraph $\mathcal{H}$ with 8 nodes and 5 hyperedges, where each node has an attribute vector and hyperedges $e_1,e_2$ contain 4 and 3 nodes, i.e., $\{v_1,v_2,v_4,v_5\}$ and $\{v_1,v_3,v_4\}$, respectively. 

\revision{
\vspace{0.5pt}\noindent \textbf{Attributed Graph} is denoted by $\mathcal{G}=(\mathcal{V}, \mathcal{E}, \XM )$, where every edge in $\mathcal{E}$ connects exactly two nodes.}
A graph $\mathcal{G}$ can be undirected or directed. An undirected edge can be viewed as two directed edges of the same node pair in reversed directions.
Different from a hypergraph incident matrix 
between nodes and hyperedges, graph adjacency matrix ${\AM} \in \mathbb{R}^{n\times n}$ encodes the structure of $\mathcal{G}$, where entry ${\AM}[i,j]$ is $1$ if there is an edge from node $v_i$ to node $v_j$, i.e., $(v_{i},v_{j}) \in \mathcal{E}_G$, or $0$ if otherwise. Let ${\DM}\in \mathbb{R}^{n\times n}$ be the diagonal node degree matrix of $\mathcal{G}$.

\revision{
\vspace{0.5pt}\noindent \textbf{Attributed Multiplex Graph} is $\mathcal{G}_M=(\mathcal{V}, \mathcal{E}_1,..., \mathcal{E}_L ,\XM )$, consisting of $L$ graph layers. Every $l$-th layer has its own edge set $\mathcal{E}_l$, and  can be viewed as an attributed graph $\mathcal{G}_l$ with $\mathcal{E}_l$, adjacency matrix $\AM_l$, and diagonal node degree matrix ${\DM}_l$.
}

\extension{\stitle{The Clustering Problem}
Given an attributed network $\mathcal{N}$ that can be  $\mathcal{H}$, $\mathcal{G}$, or $\mathcal{G}_M$, we study the clustering problem that encompasses \textit{attributed hypergraph clustering} (AHC), \textit{attributed graph clustering} (AGC), and \textit{attributed multiplex graph clustering} (AMGC).
Given a specified number $k$ of clusters and an attributed network $\mathcal{N}$, the clustering task is to divide the node set $\mathcal{V}$ into $k$ disjoint subsets $\{\mathcal{C}_1, \dots, \mathcal{C}_k\}$ such that $\bigcup_{i=1}^k \mathcal{C}_i=\mathcal{V}$ and the following properties are satisfied:}
\vspace{-\topsep}
\begin{enumerate}[leftmargin=*]
    \item Nodes within the same cluster are closely connected to each other in the network structure, while nodes in different clusters are far apart ({\bf structure closeness});
    \item Nodes in the same cluster have similar attribute values, while nodes in different clusters vary significantly in attribute values ({\bf attribute homogeneity}).
\end{enumerate}

For instance, when the input network $\mathcal{N}$ is the attributed hypergraph $\mathcal{H}$ in Figure \ref{fig:ahg}, $\mathcal{H}$ is partitioned into two clusters $\CS_1$ and $\CS_2$. We can observe that nodes $v_1$-$v_5$ in $\CS_1$ share similar attributes and are closely connected to each other, whereas nodes $v_6,v_7$ and $v_8$ form a cluster $\CS_2$ that is separated from $\CS_1$ with a paucity of connections and distinct attributes.

%% file: methodology.tex
\section{Attributed Hypergraph Clustering}\label{sec:AuHC}

\begin{figure}[!t]
    \centering
    \includegraphics[width=0.8\columnwidth]{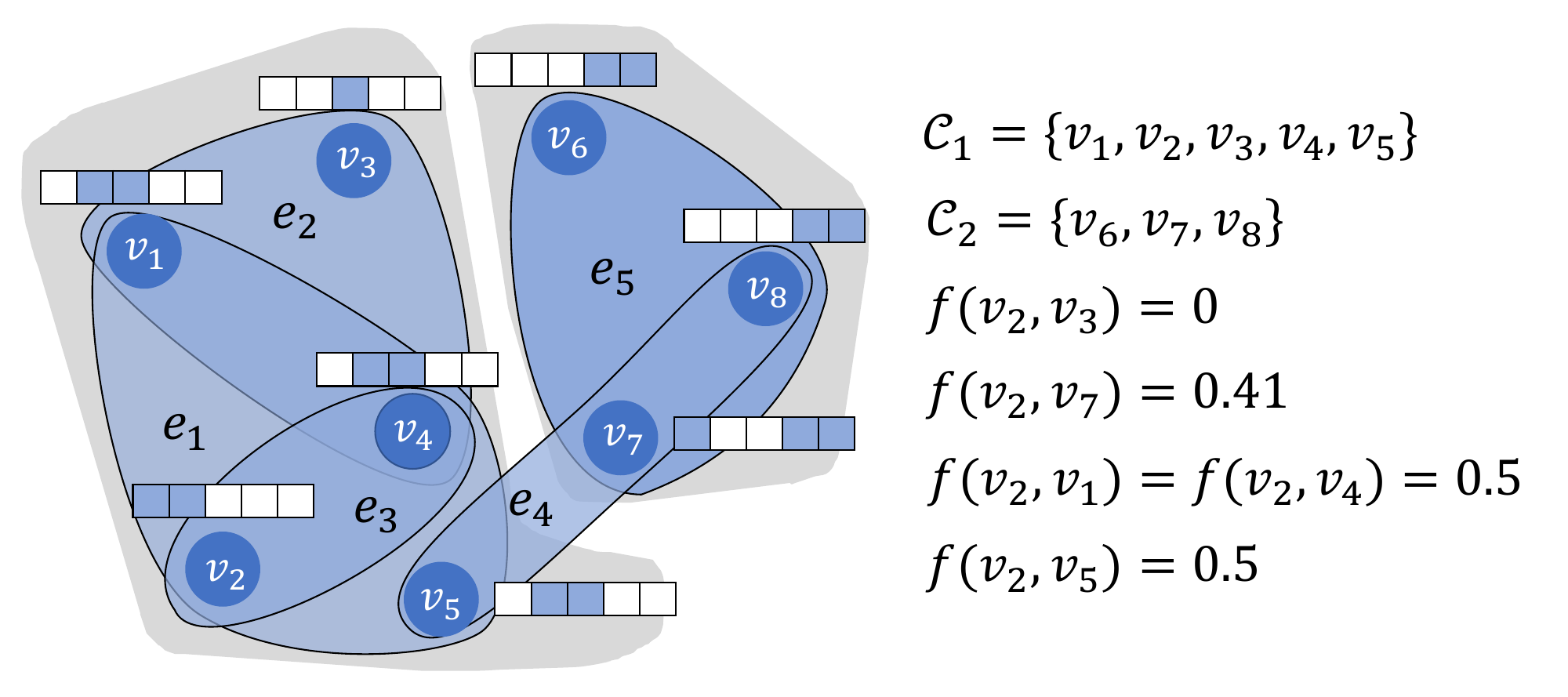}
    \vspace{-3mm}
    \caption{An Example Attributed Hypergraph}
    \label{fig:ahg}
    \vspace{-4mm}
\end{figure}

\extension{As mentioned, we first focus on attributed hypergraph clustering (AHC) and present our method \mainalgo \cite{Li2023EfficientAE} in Sections \ref{sec:AuHC}, \ref{sec:objective}, and \ref{sec:algo}.
Specifically, we will devise a random walk scheme on a K-nearest neighbor augmented hypergraph and present the AHC objective in Section \ref{sec:AuHC}, conduct theoretical analysis to support the design of \mainalgo in Section \ref{sec:objective}, and develop the algorithmic details of \mainalgo in Section \ref{sec:algo}.}   

For the problem of AHC, a central challenge is how to simultaneously exploit both hypergraph structure and attribute information for improved clustering quality.
In literature, it is a natural and effective approach to augment network structure with attribute similarity strengths
\cite{yangEffectiveScalableClustering2021,cheng2011clustering}. 
However, since a hypergraph yields different topological characteristics as illustrated in Figure \ref{fig:ahg}, we argue that attribute augmentation should be conducted in a \textit{controlable} way; otherwise, attributes may hamper, instead of improving, clustering quality, as shown in experiments (Section \ref{sec:param-exp}).

Therefore, in this section, we first develop a carefully-crafted augmentation strategy to augment attributes of nodes with hypergraph topology, which will benefit the clustering quality shown later on. \extension{As this augmentation strategy is orthogonal to the topological nature of hypergraph, its application to other types of networks, such as attributed graphs and attributed multiplex graphs, will be explained shortly in Section \ref{sec:extendagc}.}
Then we formulate Attributed Hypergraph Clustering as \textit{Augmented Hypergraph Clustering}, with the same abbreviation AHC.
The augmented hypergraph involves both hypergraph connections as well as augmented attribute connections.
It is challenging to define a unified way to preserve the high-order information of both sides. 
To tackle this, we design the {\em $(\alpha,\beta,\gamma)$-random walk} to uniformly model the node relationships (in terms of both the structural closeness and attribute similarity) in the augmented hypergraphs.
Based thereon, we define a multi-hop conductance (MHC), and formulate the objective of AHC as optimizing the conductance.

\begin{figure}[!t]
    \centering
     \scalebox{0.8}{\input{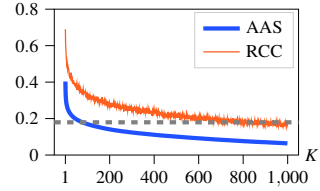}}
     \vspace{-3mm}
  \captionof{figure}{AAS and RCC on Cora-CA (best viewed in color)}
    \label{fig:sim-dist}
    \vspace{-4mm}
\end{figure}

\subsection{KNN Augmentation}\label{sec:aug-graph}

Although the vanilla augmentation strategy improves the clustering quality in attributed graphs \cite{yangEffectiveScalableClustering2021,cheng2011clustering}, to our knowledge, its effectiveness over attributed hypergraphs is as of yet under-explored. Moreover, it requires constructing a densely connected graph, causing severe efficiency issues on large graphs. 
To this end, we first demystify the attribute homogeneity of nodes within the same cluster through an empirical study on a real-world attributed hypergraph, i.e., the Cora-CA dataset\footnote{\url{https://people.cs.umass.edu/~mccallum/data.html}}
containing 2.7k academic papers in 7 research fields (i.e., 7 clusters).
\revision{Every node has an attribute vector indicating the presence of words in the corresponding publication.}
First, we use $f(v_i,v_j)=cosine(\XM[i],\XM[j])$ to denote the attribute similarity of nodes $v_i$, $v_j$. We refer to $v_j$ as the $K$-th nearest neighbor of $v_i$ if $f(v_i,v_j)$ is the $K$-th largest $\forall{v_j\in \mathcal{V}\setminus v_i}$.
Figure \ref{fig:sim-dist} plots the {\em averaged attribute similarity} (AAS for short) $f(v_i,v_j)$ of any randomly picked node $v_i$ and its $K$-th nearest neighbor $v_j$, and their ratio of co-occurring in the same cluster (RCC for short), when varying $K$ from 1 to 1000. {The AAS and RCC results from this real-world example demonstrate that two nodes with higher attribute similarity are also more likely to appear in the same cluster.} Intuitively, applying the attribute-based augmentation strategy to hypergraphs can enhance the clustering results. 

However, excessively augmenting the hypergraph with attribute information, namely, building too many connections between nodes according to attributes, will introduce distortion and adversely impact the clustering performance. To illustrate this, consider the example in Figure \ref{fig:ahg}, where nodes $v_2$, $v_3$ are in the same cluster as they share multiple common neighbors while $v_2$, $v_7$ are not. If we were to assign a cluster to node $v_2$ as per the additional connections created by attribute similarities, it is more likely to be $v_2$, $v_7$ rather than $v_2$, $v_3$ in the same cluster given $f(v_2,v_7)=0.41> f(v_2,v_3)=0$, which is counter-intuitive.

Therefore, unlike the vanilla augmentation strategy employed in prior works, we propose a KNN augmentation strategy. That is, given the input attributed hypergraph $\mathcal{H}=(\mathcal{V}, \mathcal{E}, \XM)$ and an integer $K$, we augment $\mathcal{H}$ with an undirected KNN graph $\mathcal{G}_K=(\mathcal{V},\mathcal{E}_K)$. More specifically, for each node $v_i\in \mathcal{V}$, we identify $K$ nodes in $\mathcal{V}$ (excluding $v_i$ itself) that are most similar to $v_i$ in terms of attribute similarity computed based on a similarity function $f(\cdot,\cdot)$ as $v_i$'s neighbors in $\mathcal{G}_K$, denoted by $N_K(v_i)$. In other words, for every two nodes $v_i$, $v_j$ ($v_j\in N_K(v_i)$), we construct an edge $(v_i,v_j)$ with weight $f(\XM[i],\XM[j])$ in $\mathcal{E}_K$.
Accordingly, the adjacency matrix $\AM_K$ of $\mathcal{G}_K$ is defined as follows:
\vspace{-1mm}
\begin{equation}\label{AK}
\resizebox{0.9\hsize}{!}{$
    \AM_K[i, j]=\begin{cases}0, &\text{if }v_i\notin N_K(v_j) \text{ and } v_j\notin N_K(v_i),\\
    2\cdot f(\XM[i], \XM[j]), &\text{if }v_i\in N_K(v_j) \text{ and } v_j\in N_K(v_i),\\
    f(\XM[i], \XM[j]), &\text{otherwise}.
    \end{cases}$}
\end{equation}
Thus, we obtain an augmented hypergraph $\mathcal{H}_A$ containing the hypergraph $\mathcal{H}_O=(\mathcal{V},\mathcal{E})$ and the KNN graph $\mathcal{G}_K=(\mathcal{V},\mathcal{E}_K)$.
The reasons that we only consider $K$ nearest neighbors for augmented hypergraph construction are three-fold. In the first place, the case study in Figure \ref{fig:sim-dist} suggests that there is no significant difference between the RCC of two random nodes (depicted by the gray dashed line) and that of two nodes $v_i,v_j$ such that $v_j\in N_K(v_i)$, when $K$ is beyond a number (roughly $500$ in Figure \ref{fig:sim-dist}). Therefore, such connections can be overlooked without impeding the clustering quality.
Secondly, if we revisit the example in Figure \ref{fig:ahg} and apply the KNN strategy ($K=3$) here, we can exclude the connection between $v_2$ and $v_7$ from $\mathcal{G}_K$ since $f(v_2,v_1)=f(v_2,v_4)=f(v_2,v_5)=0.5> f(v_2,v_7)=0.41$. The distortion issue mentioned previously is therefore resolved.
In comparison with the densely connected graph that encodes all attribute similarities (with up to $O(n^2)$ edges in the worst case), 
$\mathcal{G}_K$ can be efficiently constructed by utilizing well-established approximate nearest neighbor techniques \revision{with $O(n\log n)$ complexity~\cite{guo_accelerating_2020,johnson2019billion}.}

{The range of the KNN neighborhood is determined by parameter $K$. While a larger $K$ allows the KNN graph to include more attribute similarity relations, this also leads to a higher proportion of unwanted inter-cluster edges in the KNN graph as evidenced by the lower RCC in Figure \ref{fig:sim-dist}. 
Meanwhile, $K$ cannot be too small (e.g., 5), or it will fail to utilize highly similar nodes that usually have high RCC.
The trade-off of choosing $K$ is evaluated in Section \ref{sec:param-exp}.}

Now, the question lies in how to model the relationships of nodes in $\mathcal{V}$ of the augmented graph $\mathcal{H}_A$, which is a linchpin to AHC.
In the following section, we present a joint random walk model that enables us to capture the multi-hop proximities of nodes over $\mathcal{H}_O$ and $\mathcal{G}_K$ jointly.

\subsection{$(\alpha,\beta,\gamma)$-Random Walk}\label{sec:rand-walk}

Random walk with restart \cite{tong2006rwrapplication} (RWR) is one of the most common and effective models for capturing the multi-hop relationships between nodes in a graph \cite{jung2017bepirandomwalk}, and is widely used in many tasks such as ranking \cite{tong2006rwrapplication,ShiYJXY19}, recommendation \cite{park2017rwrrecommendation}, and clustering \cite{pmlr-v28-allenzhu13}. Given a graph $\mathcal{G}$, a source node $u$ and a stopping probability $\alpha$ (typically $\alpha=0.2$), at each step, an RWR originating from $u$ either stops at the current node with probability $\alpha$, or randomly picks an out-neighbor $v$ of the current node according to the weight of edge $(u,v)$ and navigates to $v$ with the remaining $1-\alpha$ probability. It follows that RWR score (a.k.a. {\em personalized PageRank} \cite{jeh2003scaling}) of any node pair $(u,v)$ represents the probability that an RWR from $u$ ends at node $v$. Intuitively, two nodes with dense (one-hop or multi-hop) connections should have a high RWR score.

Nevertheless, RWR is designed for general graphs, and thus cannot be directly applied to our augmented hypergraph $\mathcal{H}_A$ as it consists of a hypergraph $\mathcal{H}_O$ and a general graph $\mathcal{G}_K$. We devise a joint random walk scheme, named $(\alpha,\beta,\gamma)$-random walk, which conducts the RWR process over $\mathcal{H}_O$ and $\mathcal{G}_K$ jointly to seamlessly integrate topological proximity over both networks.
Definition \ref{def:rw-aug-hg} states the formal definition of the $(\alpha,\beta,\gamma)$-random walk process.
\begin{definition} \label{def:rw-aug-hg}
Given an augmented hypergraph $\mathcal{H}_A=(\mathcal{H}_O,\mathcal{G}_K)$ and a source node $u$, an $(\alpha,\beta,\gamma)$-random walk $W$ starting from $u$ conducts $\gamma$ steps and at each step proceeds as follows.
\vspace{-\topsep}
\begin{itemize}[leftmargin=*]
\item With probability $\alpha$, $W$ terminates at the current node $v_i$;
\item with the other $1-\alpha$ probability, $W$ navigates to a node $v_j$ picked by the following rules:
\begin{itemize}
    \item with probability $\beta_i$, $W$ draws an out-neighbor $v_j$ of the current node $v_i$ in $\mathcal{G}_K$ according to probability $\frac{\AM_K[i,j]}{\sum_{v_l\in N_K(v_i)}{\AM_K[i,l]}}$;
    \item or with probability $1-\beta_i$, $W$ first draws an hyperedge $e_i$ incident to $v_i$ in $\mathcal{H}_O$, and then draws node $v_j$ from $e_i$ uniformly at random.
\end{itemize}
\end{itemize}

\end{definition}
Each node $v_i$ is associated with a parameter $\beta_i$ (see Eq. \eqref{eq:beta_i}) used to control the joint navigation between hypergraph $\mathcal{H}_O$ and KNN $\mathcal{G}_K$.
The larger $\beta_i$ is, the more likely that the random walk jumps to the neighbors of $v_i$ in KNN $\mathcal{G}_K$.
\begin{equation}\label{eq:beta_i}
    \beta_i = \begin{cases}
        0, & \text{if } \XM[i] \textrm{\ is a zero vector};\\
        1, & \text{else if } \delta(v_i)=0;\\
        \beta, & \text{otherwise}.
    \end{cases}
\end{equation}
In general, we set $\beta_i$ to $\beta\in [0,1]$, which is a user-specified parameter.
In particular cases, when node $v_i$'s attribute vector $\XM[i]$ is a zero vector, i.e., $v_i$ has no useful information in the KNN $\mathcal{G}_K$, we set $\beta_i$ to $0$. Conversely, $\beta_i$ is configured as $1$ if $v_i$ is connected to none of the hyperedges, i.e., $\delta(v_i)=0$. 
Let $s(v_i,v_j)$ denote the probability of an $(\alpha,\beta,\gamma)$-random walk from $v_i$ stopping at $v_j$ in the end.
Based on Definition \ref{def:rw-aug-hg}, we can derive the following formula for $s(v_i,v_j)$:
\begin{equation}\label{eq:approx-multi-hop}
\begin{aligned}
    s(v_i,v_j)=\SM[i,j] &\textstyle = \alpha \sum_{\ell=0}^{\gamma} (1-\alpha)^{\ell}\PM^{\ell}[i,j], \\
\end{aligned}
\end{equation}
where $\PM$ is a transition matrix defined by
\begin{equation}\label{eq:p-mat}
    \PM = (\IM-\BM)\cdot\DM_V^{-1} \HM\transpose \DM_E^{-1} \HM + \BM \DM_K^{-1} \AM_K,
\end{equation}
$\BM=diag(\beta_1, \dots, \beta_n)$ is a diagonal matrix containing $\beta_i$ parameters, and $\DM_K$ is the diagonal degree matrix of   $\mathcal{G}_K$.  $\PM^{\ell}[i,j]$ is the probability that a $\ell$-hop walk from $v_i$ terminates at $v_j$.

\subsection{Objective Function}\label{sec:objfunc}
In what follows, we formally define the objective function of AHC. Intuitively, a high-quality cluster $\CS$ in the augmented hypergraph $\mathcal{H}_A$ should be both internally cohesive and well disconnected from the remainder of the graph with the consideration of multi-hop connections. Hence, if we simulate an $(\alpha,\beta,\gamma)$-random walk $W$ from any node in $\CS$, $W$ should have a low probability of escaping from $\CS$, i.e., ending at any node outside $\CS$.
We refer to this escaping probability $\phi(\CS)$ as the {\em multi-hop conductance} (MHC) of $\CS$, defined in Eq. \eqref{eq:phi-c}. 
\begin{equation}\label{eq:phi-c}
 \textstyle   \phi(\CS) = \frac{1}{|\CS|} \sum_{v_i \in \CS} \sum_{v_j\notin \CS} s(v_i, v_j)
\end{equation}
Since a low MHC $\phi(\CS)$ reflects a high coherence of cluster $\CS$, we then formulate AHC as an optimization problem of finding $k$ clusters $\{\CS_1,\dots,\CS_k\}$ such that their MHC $\Phi(\{\CS_1, \dots, \CS_k\})$ (Eq. \eqref{eq:mh-cond}) is minimized.
\begin{equation}\label{eq:mh-cond}
    \Phi(\{\CS_1, \dots, \CS_k\}) = \frac{1}{k} \sum_{\CS\in\{\CS_1,\dots,\CS_k\}} \frac{1}{|\CS|} \sum_{v_i \in \CS} \sum_{v_j\notin \CS} s(v_i, v_j)
\end{equation}

Directly minimizing Eq. \eqref{eq:mh-cond} requires computing $s(v_i,v_j)$ (Eq. \eqref{eq:approx-multi-hop}) of every two nodes $v_i\in \CS$, $v_j\in \mathcal{V}\setminus \CS$, $\forall{\CS}\in \{\CS_1,\CS_2,\\\cdots,\CS_k\}$, which is prohibitively expensive due to intractable computation time (i.e., $O(n^3)$) and storage space (i.e., $O(n^2)$). In addition, the minimization of $\Phi(\{\CS_1, \dots, \CS_k\})$ is an NP-complete combinatorial optimization problem \cite{shi2000normalizedcuttpami}, rendering the exact solution unattainable on large graphs.

\section{Theoretical Analysis for \mainalgo}\label{sec:objective}

This section presents the top-level idea of our proposed solution, \mainalgo, to AHC computation, and explains the intuitions behind it. 
At a high level, \mainalgo first transforms the objective of AHC in Eq. \eqref{eq:mh-cond} to a matrix trace maximization problem, and then derives an approximate solution via a top-$k$ eigendecomposition.
Note that for any $k$ non-overlapping clusters $\{\CS_1,\CS_2,\cdots,\CS_k\}$ on $\mathcal{H}$ satisfying $\bigcup_{i=1}^k \mathcal{C}_i=\mathcal{V}$, they can be represented by a binary matrix ${\YM}\in \{0,1\}^{n\times k}$, where for each node $v_i$ and cluster $\CS_j$
\begin{equation}\label{eq:y}
    {\YM}[i, j] = \begin{cases}
        1, & v_i\in \CS_j\\
        0, & v_i\in \mathcal{V}\setminus \CS_j.
    \end{cases}
\end{equation}
We refer to ${\YM}$ as a {\em binary cluster membership} (BCM) matrix of $\mathcal{H}$ and we use 
\begin{equation}\label{eq:norm-y}
h(\YM) = ({\YM}\transpose {\YM})^{-1/2}{\YM} = \widehat{\YM}  
\end{equation}
to stand for the $L_2$ normalization of $\YM$. Particularly, $\widehat{\YM}$ has orthonormal columns, i.e., $\widehat{\YM}\transpose\widehat{\YM}=\IM_k$ where $\IM_k$ is a $k\times k$ identity matrix.
Given $k$ non-overlapping clusters $\{\CS_1,\CS_2,\cdots,\CS_k\}$ and their corresponding BCM matrix $\YM$, it is trivial to show
\begin{equation}\label{eq:trace-cond}
        \Phi(\{\CS_1,\dots,\CS_k\})= 1-\Psi(\YM),
    \end{equation}
    where $\Psi(\YM)$ is defined as follows:
\begin{equation}\label{eq:psi-y}
\Psi(\YM)=\frac{1}{k}trace(\widehat{\YM}\transpose \SM \widehat{\YM}).
\end{equation}
Eq. \eqref{eq:trace-cond} suggests that the minimization of MHC $\Phi(\{\CS_1,\dots,\CS_k\})$ is equivalent to finding a BCM matrix $\YM$ such that the trace of matrix $\widehat{\YM}\transpose \SM \widehat{\YM}$ is maximized. Due to its NP-completeness, instead of computing the exact solution, we utilize a two-phase strategy to derive an approximate solution as follows.

If we relax the binary constraint on $\YM$, the following lemma establishes an upper bound $\psi_\sigma$ for $\Psi(\YM)$.
\begin{lemma}\label{lemma:sigma-lb}
Let $\sigma_1\geq \sigma_2\geq\dots\geq\sigma_k$ be the $k$ largest singular values of matrix $\SM$ in Eq. \eqref{eq:approx-multi-hop}. Given any matrix $\WM\in \mathbb{R}^{n\times k }$ such that $h(\WM)$ satisfies ${h(\WM)}^\top \cdot h(\WM)=\IM_K$, then
        $\Psi(\WM) \le \frac{1}{k}\sum_{i=1}^{k}\sigma_i = \psi_\sigma$.
\end{lemma}

Lemma \ref{lemma:sigma-lb} implies that if we can first find a fractional matrix $\WM$ such that $\Psi(\WM)$ is close to $\psi_\sigma$, a high-quality BCM matrix $\YM$ can be converted from $\WM$ by leveraging algorithms such as $k\textrm{-}\mathtt{Means}$~\cite{lloydLeastSquaresQuantization1982}.
Although we can obtain such a fractional matrix $\WM$ by applying trace maximization techniques \cite{won2021maxtrace} to Eq. \eqref{eq:psi-y}, it still remains tenaciously challenging to compute $\SM$. (\inPaper{All proofs are in the technical report \cite{report}.})

\begin{lemma}\label{lemma:lambda-lb}
Let the columns of $\QM\in \mathbb{R}^{n\times k}$ be the second to $(k+1)$-th leading eigenvectors of $\PM$ (Eq. \eqref{eq:p-mat}). Then, we have
        $\Psi(\QM) = \frac{1}{k}\sum_{i=2}^{k+1}\lambda_i = \psi_\lambda$,
where $\lambda_2\geq\dots\geq\lambda_k\geq \lambda_{k+1}$ are the second to $(k+1)$-th leading eigenvalues of $\SM$, sorted by algebraic value in descending order.
\end{lemma}

We exclude the first eigenvector $\frac{1}{\sqrt{n}}\cdot \mathbf{1}$ of $\PM$ as it is useless for clustering. \revision{By virtue of our analysis in Lemma \ref{lemma:lambda-lb}, the second to $(k+1)$-th leading eigenvectors $\QM$ of $\PM$ (see Eq. \eqref{eq:p-mat}) can be regarded as a rough $\WM$ since $\Psi(\QM)=\psi_\lambda \le  \psi_\sigma$ and the gap between $\psi_\lambda$ and  $\psi_\sigma$ is insignificant in practice.} 
For instance, on the Cora-CA dataset, we can obtain $\psi_\sigma = 0.668$ and $\psi_\lambda = 0.596$ \revision{(\ie $\Phi_\sigma=1-\psi_\sigma=0.332$, $\Phi_\lambda=1-\psi_\lambda=0.404$)}, both of which are better than  $\Psi(\YM^{\ast})=0.533$ \revision{(\ie $\Phi^{\ast}=1-\Psi(\YM^{\ast})=0.467$)} of the ground-truth BCM matrix $\YM^{\ast}$. Consequently, using the second to $(k+1)$-th leading eigenvectors $\QM$ of $\PM$ as the fractional solution $\WM$ is sufficient to derive a favorable BCM matrix. Moreover, in doing so, we can avoid the tremendous overhead incurred by the materialization of $\SM$. 

To summarize, \mainalgo adopts a two-phase strategy to obtain an approximate solution to the AHC problem. First, \mainalgo computes the second to $(k+1)$-th leading eigenvectors $\QM$ of $\PM$. After that, \mainalgo transforms $\QM$ into a BCM matrix $\YM$ through a discretization approach \cite{yu_multiclass_2003} that minimizes the difference between $\QM$ and $\YM$. The rationale is that $\Psi(\QM)=\Psi(\QM\RM)$ if $\RM$ is a $k \times k$ orthogonal matrix, ensuring $\RM\transpose\RM=\IM_k$. Accordingly, we can derive a BCM matrix $\YM=\QM\RM$ by minimizing the Frobenius norm $||\QM - \QM\RM||_F$ with a binary constraint exerted on $\QM\RM$. Note that we do not adopt $k\textrm{-}\mathtt{Means}$ over $\QM$ to get the BCM matrix $\YM$ as it deviates from the objective in Eq. \eqref{eq:psi-y}, and thus, produces sub-par result quality, as revealed by experiments (Table \ref{tab:ablation}).

Nevertheless, to realize the above idea, there still remain two crucial technical issues to be addressed:
\vspace{-\topsep}
\begin{enumerate}[leftmargin=*]
\item The brute-force computation of $\QM$ is time-consuming as it requires numerous iterations and the construction of $\PM$.
\item In practice, directly utilizing the exact or near-exact $\QM$ might incur overfitting towards the objective instead of ground-truth clusters, and hence, lead to sub-optimal clustering quality. It is challenging to derive a practically effective and robust BCM matrix $\YM$ from $\QM$.
\vspace{-3mm}
\end{enumerate}

\begin{figure}[!t]
    \centering
    \resizebox{\columnwidth}{!}{
        \input{fig-overview.tikz}
    }
    \vspace{-4mm}
    \caption{Overview of \mainalgo}
    \label{fig:overview}
    \vspace{-3mm}
\end{figure}

\section{The \mainalgo Algorithm}\label{sec:algo}

To circumvent the above challenges, \mainalgo integrates the aforementioned two-phase scheme into an iterative framework, which enables us to approximate the second to $(k+1)$-th leading eigenvectors $\QM$ without constructing $\PM$ explicitly, and greedily search the BCM matrix $\YM$ with the best MHC.
Figure \ref{fig:overview} sketches the main ingredients and algorithmic procedure of \mainalgo. More specifically, \mainalgo employs {\em orthogonal iterations} \cite{saad1992numerical} to approximate the second to $(k+1)$-th leading eigenvectors $\QM$ of $\PM$. During the course, \mainalgo starts with an initial BCM matrix, followed by an orthogonal iteration to compute an approximate $\QM$ and an updated BCM matrix $\YM$ from the $\QM$ through $\mathtt{Discretize}$ algorithm \cite{yu_multiclass_2003}. 
Afterward, \mainalgo inspects if $\QM$ reaches convergence and computes the MHC with the current BCM matrix $\YM$ via \mhc{} algorithm (Algorithm \ref{alg:mhcond}). If $\QM$ converges (i.e., the BCM remains nearly stationary 
) 
or the early termination condition is satisfied (i.e., the MHC of current $\YM$ is satisfying), \mainalgo terminates. Otherwise, \mainalgo enters into the next orthogonal iteration with the updated $\QM$ and $\YM$.

In what follows, a detailed description of \mainalgo is given in Section \ref{sec:main-algo}.
Section \ref{sec:init} introduces 
an effective approach \bcm{} for initializing the BCM matrix $\YM$, which drastically curtails the number of iterations needed and significantly boosts the computation efficiency of \mainalgo. The complexity of the complete algorithm is analyzed in Section \ref{sec:analysis}.

\vspace{-1mm}
\subsection{Main Algorithm}\label{sec:main-algo}
\vspace{-1mm}

\begin{algorithm}[!t]
\caption{\mainalgo{}}\label{alg:mainalg}
\KwIn{Hypergraph $\mathcal{H}$, KNN transition matrix $\PM_K$, the number of clusters $k$, diagonal matrix $\BM$, constant $\alpha$,  error threshold $\epsilon_Q$, the numbers of iterations $T_a$, $\gamma$, an integer $\tau$, and an initial BCM matrix ${\YM}^{(0)}$.}
\KwOut{BCM matrix $\YM$}
 $\YM \gets \YM^{(0)}$, $ \widehat{\YM}^{(0)} \gets h({\YM}^{(0)})$\;
 $\QM^{(0)}\gets \frac{\mathbf{1}}{\sqrt{n}}\cdot \mathbf{1} | \widehat{\YM}^{(0)}$ \;
 \For{$t \gets 1, 2, \cdots, T_a$}{
  Compute $\ZM^{(t)}$ according to Eq. \eqref{eq:zt}\;
  $\QM^{(t)}, \RM^{(t)}\gets \texttt{QR}(\ZM^{(t)})$ \;
  \If{t $mod\ \tau = 0$ }{
   ${\YM}^{(t)} \gets $ \texttt{Discretize}($\QM^{(t)}$) \;
   {
   $\Phi({\YM}^{(t)}) \gets \mhc{}({\YM}^{(t)}, \PM_V, \PM_E, \PM_K, \BM, \gamma, \alpha)$\;
   \lIf{$\Phi({\YM}^{(t)})<\Phi({\YM})$}{$\YM \gets \YM^{(t)}$}
   \lIf{Eq. \eqref{eq:tcond-1} or Eq. \eqref{eq:tcond-2} holds}{\textbf{break}} 
   }
  }
 }
 {
 \Return ${\YM}$\;
 }
\end{algorithm}

The pseudo-code of \mainalgo is presented in Algorithm \ref{alg:mainalg}, which takes as input an attributed hypergraph $\mathcal{H}$, transition matrix of attribute KNN graph $\PM_K$, the number $k$ of clusters, a diagonal matrix $\BM$ containing $n$ parameters defined in Eq. \eqref{eq:beta_i}, the random walk stopping probability $\alpha$, an error threshold $\epsilon_Q$, the numbers $\gamma, T_a$ of iterations, an integer $\tau$, and an initial BCM matrix ${\YM}^{(0)}$. \mainalgo starts by computing the normalized BCM matrix $\widehat{\YM}^{(0)}= h(\YM^{(0)})$ (Eq. \eqref{eq:norm-y}) and setting the initial $k+1$ leading eigenvectors $\QM^{(0)}$ as $\frac{1}{\sqrt{n}}\cdot\mathbf{1} | \widehat{\YM}^{(0)}$ (Lines 1-2), where $|$ represents the horizontal concatenation and $\frac{1}{\sqrt{n}}\cdot\mathbf{1}$ is the first leading eigenvector of $\PM$ since it is a stochastic matrix. 
After that, \mainalgo enters into at most $T_a$ orthogonal iterations for computing the $k+1$ leading eigenvectors $\QM$ and the BCM matrix $\YM$ (Lines 3-10). 
At step $t$, orthogonal iteration updates the approximate $k+1$ leading eigenvectors of $\PM$ as $\QM^{(t)}$ by the formula below (Lines 4-5):
\begin{equation}\label{eq:ortho-iter}
    \QM^{(t)}\RM^{(t)}=\ZM^{(t)}=\PM \QM^{(t-1)},
\end{equation}
where $\QM^{(t)}$ is obtained by a QR decomposition over $\ZM^{(t)}$. If $t$ is sufficiently large, $\QM^{(t)}$ will converge to the exact $k+1$ leading eigenvectors of $\PM$ \cite{saad1992numerical}.
Note that the direct computation of $\ZM^{(t)}=\PM\QM^{t-1}$ requires constructing $\PM$ explicitly as per Eq. \eqref{eq:p-mat}, which incurs an exorbitant amount of time and space (up to $O(n^2)$ in the worst case).
To mitigate this, we decouple and reorder the matrix multiplication as in Eq. \eqref{eq:zt}.
\begin{align}
&\ZM^{(t)} = (\IM-\BM)\cdot\PM_V \cdot \left(\PM_E \QM^{(t-1)}\right) + \BM\PM_K\cdot \QM^{(t-1)}, \label{eq:zt}\\
&\text{where } \PM_V=\DM^{-1}_V\HM\transpose,\ \PM_E=\DM^{-1}_E\HM\label{eq:pv-pe}
\end{align}
$\PM_V$ and $\PM_E$ are two sparse matrices of $\mathcal{H}$ and $\PM_K=\DM_K^{-1} \AM_K$ is the sparse transition matrix of the KNN graph $\mathcal{G}_K$ defined in Section \ref{sec:aug-graph}.
Note that all of them can be efficiently constructed in the preprocessing stage. 
As such, we eliminate the need to materialize $\PM$ and reduce the time complexity of computing $\ZM^{(t)}$ to $O(nk\cdot (\overline{\delta} +K))$. 

\begin{algorithm}[!t]
\caption{\mhc{}}\label{alg:mhcond}
\KwIn{${\YM}^{(t)}, \PM_V, \PM_E, \PM_K, \BM, \gamma, \alpha$}
\KwOut{MHC $\phi_t$}
 $\widehat{\YM}^{(t)} \gets h(\YM^{(t)});\ \FM^{(0)} \gets \alpha\widehat{\YM}^{(t)}$\;
 \For{$\ell \gets 1, 2, \dots \gamma$}{
  Compute $\FM^{(\ell)}$ according to Eq. \eqref{eq:mhc-comp};
 }
 $\phi_t \gets 1-\frac{1}{k}trace(\widehat{\YM}^{(t)\top} \FM^{(\gamma)})$\;
 \Return $\phi_t$ \;
\end{algorithm}

After obtaining $\QM^{(t)}$, \mainalgo converts $\QM^{(t)}$ into a new BCM matrix $\YM^{(t)}$ (Lines 6-7) using the \texttt{Discretize} algorithm~\cite{yu_multiclass_2003}. Notice that we conduct this conversion every other $\tau$ iterations in order to avert unnecessary operations as the difference between $\YM^{(t)}$ and $\YM^{(t-1)}$ is often insignificant.

Next, at Line 8, \mainalgo invokes \mhc{} (i.e., Algorithm \ref{alg:mhcond}) with a BCM matrix ${\YM}^{(t)}$, other parameters including $\PM_V$, $\PM_E$, $\PM_K$, $\BM$, $\alpha$, and the number of iterations $\gamma$ as input to calculate the MHC $\phi_t$ of the current BCM matrix $\YM^{(t)}$. To avoid the materialization of $\SM$ required in Eq. \eqref{eq:trace-cond} and Eq. \eqref{eq:psi-y}, Algorithm \ref{alg:mhcond} computes $\phi_t$ in an iterative manner by reordering the matrix multiplications (Lines 2-3 in Algorithm \ref{alg:mhcond}). More precisely, at the $\ell$-th iteration, it obtains the intermediate result $\FM^{(\ell)}$ via the following equation: 
\begin{equation}\label{eq:mhc-comp}
\resizebox{0.9\hsize}{!}{$
\textstyle \FM^{(\ell)} = (1-\alpha) \left((\IM-\BM)\cdot \PM_V \cdot \left(\PM_E \FM^{(\ell-1)}\right) + \BM \PM_K\cdot \FM^{(\ell-1)}\right) + \FM^{(0)}.$}
\end{equation}
$\FM^{(0)}$ is initialized as Line 1 in Algorithm \ref{alg:mhcond}. It can be verified that $\phi_t=1-\frac{1}{k}trace({\widehat{\YM}^{(t)\top}} \FM^{(\gamma)})$ (Line 4 in Algorithm \ref{alg:mhcond}).

Once the convergence criterion of $\QM^{(t)}$ (Eq. \eqref{eq:tcond-1}) is satisfied, or the early termination condition (Eq. \eqref{eq:tcond-2}) holds, \mainalgo ceases the iterative process and {returns the BCM matrix $\YM$ with the lowest MHC (Lines 9-11 in Algorithm \ref{alg:mainalg}). }
\begin{align}
    &||\QM^{(t)}-\QM^{(t-1)}||<\epsilon_Q \label{eq:tcond-1} \\
    &\phi_{t-2\tau}<\phi_{t-\tau}<\phi_{t} \label{eq:tcond-2}
\end{align}
Otherwise, \mainalgo proceeds to the next orthogonal iteration. The rationale for the early termination condition in Eq. \eqref{eq:tcond-2} \revision{is that, in practice, successive increases in $\phi_t$ indicate that clusters with desirable MHC objective have been attained.}

\subsection{Greedy Initialization of BCM}\label{sec:init}

Akin to many optimization problems, \mainalgo requires many iterations to achieve convergence when $\YM^{(0)}$ is randomly initialized. To tackle this issue, we propose a greedy initialization technique, \bcm{}, whereby we can immediately gain a passable BCM matrix $\YM^{(0)}$ and expedite the convergence, as demonstrated by our experiments in Section \ref{sec:cvg-als}.

The rationale of \bcm{} is that most nodes tend to cluster together around a number of center nodes~\cite{rattigan2007medroidsicml}. 
Therefore, we can first pick a set $\mathcal{V}_c$ of top influential nodes w.r.t. the whole hypergraph, and calculate the multi-hop proximities (i.e., RWR scores) of each node to the influential nodes $\mathcal{V}_c$ (i.e., centers). Then, the cluster center of each node can be determined by its proximity to nodes in $\mathcal{V}_c$ accordingly. 

\begin{algorithm}[!t]
\caption{\bcm{}}\label{alg:init}
\KwIn{Hypergraph $\mathcal{H}$, matrices $\PM_V, \PM_E$, integer $k$, constant $\alpha$, the number of iterations $T_i$.}
\KwOut{An initial BCM matrix ${\YM}^{(0)}$.}
 $\mathcal{V}_c \gets $ The sorted indices of nodes with $k$ largest degrees\;
 Initialize $\ZM_0\gets \mathbf{0}^{k\times n}$\;
 \lFor{$j\gets 1$ to $k$}{
 $\ZM_0[j,\mathcal{V}_c[j]] \gets 1$
 }
 Initialize $\boldsymbol{\Pi}_c^{(0)} \gets \alpha \ZM_0$\;
 \For{$t \gets 1, 2, \dots T_i$}{
  Compute $\boldsymbol{\Pi}_c^{(t)}$ according to Eq. \eqref{eq:pi_c_t}\;
 }
 \For{$v_j\in \mathcal{V}$}{Calculate $g(v_j)$ according to Eq. \eqref{eq:argmaxlabel}\;
 ${\YM}^{(0)}[j,g(v_j)]\gets 1$\;
 }
 \Return ${\YM}^{(0)}$ \;
\end{algorithm}

Algorithm \ref{alg:init} displays the pseudo-code of \bcm{}. Given hypergraph $\mathcal{H}$, and transition matrices $\PM_V, \PM_E$ defined in Eq. \eqref{eq:pv-pe}, the number $k$ of clusters, random walk stopping probability $\alpha$, and the number of iterations $T_i$, as input, \bcm{} begins by initializing an ordered set $\mathcal{V}_c$ consisting of the $k$ nodes with $k$ largest degrees in $\mathcal{H}$ (sorted by their indices), which later serves as the cluster centers (Line 1).
Then, a $k\times n$ matrix $\ZM_0$ is created, where for each integer $j\in [1,k]$, $\ZM_0[j,\mathcal{V}_c[j]]$ is set to 1 and 0 otherwise and $\mathcal{V}_c[j]$ denotes the node index of the $j$-th node in $\mathcal{V}_c$ (Lines 2-3). Next, \bcm{} launches $T_i$ iterations to calculate the RWR scores of all nodes w.r.t the $k$ nodes in $\mathcal{V}_c$ (Lines 5-6). Specifically, at $t$-th iteration, we compute approximate RWR $\boldsymbol{\Pi}_c^{(t)}$ (Line 6):
\begin{equation}\label{eq:pi_c_t}
    \boldsymbol{\Pi}_c^{(t)} = (1-\alpha) \left(\boldsymbol{\Pi}_c^{(t-1)} \PM_V\right)\cdot \PM_E + \boldsymbol{\Pi}_0,
\end{equation}
where $\boldsymbol{\Pi}_0=\alpha\ZM_0$ (Line 4). Note that we reorder the matrix multiplications as in Eq. \eqref{eq:pi_c_t} so as to bypass the materialization of the $n\times n$ matrix $\PM_V\PM_E$. After obtaining $\boldsymbol{\Pi}_c^{(T_i)}$, \bcm{} assigns the node $\mathcal{V}_c[g(v_j)]$ as the cluster center to each node $v_j$ in $\mathcal{H}$ as per Eq. \eqref{eq:argmaxlabel} (Lines 7-9).
\begin{equation}\label{eq:argmaxlabel}
g(v_j) = \arg\max_{1\le l\le k} \boldsymbol{\Pi}_c^{(T_i)}[l, j], 
\end{equation}
meaning that we pick a cluster center from $\mathcal{V}_c$ such that its RWR score $\boldsymbol{\Pi}_c^{(T_i)}[l, j]$ w.r.t $v_j$ is the highest. Finally, an $n\times k$ binary matrix $\YM^{(0)}$ is constructed by setting $\YM^{(0)}[j,g(v_j)]$ to $1$ for $v_j\in \mathcal{V}$ and returned as the initial BCM matrix. 

\subsection{Complexity}\label{sec:analysis}

One of the main computational costs of \mainalgo stems from the sparse matrix multiplications, i.e., Line 4 in Algorithm \ref{alg:mainalg}, Line 3 in Algorithm \ref{alg:mhcond}, and Line 6 in Algorithm \ref{alg:init}.
We first consider Line 4 in Algorithm \ref{alg:mainalg}, i.e., Eq. \eqref{eq:zt}. Since $\QM^{(t-1)}$ is an $n \times (k+1)$ matrix and the numbers of non-zero entries in sparse matrices $\PM_V$, $\PM_E$, and $\PM_K$ are $n\overline{\delta}$, $n\overline{\delta}$, and $nK$, respectively, its complexity is $O((n\overline{\delta}+nK)\cdot k)$ \cite{yuster2005fast}. Analogously, according to Eq. \eqref{eq:mhc-comp}, and Eq. \eqref{eq:pi_c_t}, both the time costs of Line 3 in Algorithm \ref{alg:mhcond} and Line 6 in Algorithm \ref{alg:init} are bounded by $O(n\overline{\delta}k)$. Recall that these three operations are conducted up to $T_a$, $\gamma$, and $T_i$ times in Algorithms \ref{alg:mainalg}, \ref{alg:mhcond}, and \ref{alg:init}, respectively. Therefore, the total time cost of sparse matrix multiplications is $O(kn\overline{\delta}\cdot (T_a+T_i+\gamma) + knKT_a)$. Moreover, in Algorithm \ref{alg:mainalg}, the \texttt{QR} decomposition at Line 5 takes $O(k^2 n)$ time and \texttt{Discretize}~\cite{yu_multiclass_2003} runs in $O(k^2 n+k^3)$ time. Overall, the time complexity of \mainalgo is $O(kn\overline{\delta}\cdot (T_a+T_i+\gamma) + knKT_a + k^2 n)$, which equals $O(n\overline{\delta})$ when $T_a, T_i, \gamma, k$, and $K$ are regarded as constants.
The space complexity of \mainalgo is $O(n\cdot (\overline{\delta}+K+k))$ as all matrices are in sparse form.

%% file: extension.tex
\section{\extension{The \extendagc framework}}\label{sec:extendagc}

\extension{In this section, we generalize \mainalgo that is for AHC to a versatile framework \extendagc to process all of AHC, AGC, and AMGC, formulated in Section \ref{sec:pre}. \extendagc aims to efficiently find high-quality clusters on various types of network $\mathcal{N}$.}

\extension{As mentioned, the proposed KNN augmentation in Section \ref{sec:aug-graph} is orthogonal to the high-order nature of hypergraph, and therefore, we can apply the KNN augmentation to input attributed network $\mathcal{N}$ that can be an attributed hypergraph $\mathcal{H}$, graph $\mathcal{G}$, and multiplex graph $\mathcal{G}_M$.} 

\extension{Recall that, in Figure \ref{fig:sim-dist}, we have empirically shown that nodes with higher attribute similarity are more likely to appear in the same cluster of a hypergraph $\mathcal{H}$. {This also holds for} attributed graphs and attributed multiplex graphs.
\revision{Figures \ref{fig:sim-dist-add1}-\ref{fig:sim-dist-add2} illustrate the AAS and RCC on the attributed graph Citeseer-DG and the attributed multiplex graph ACM, with binary keyword vectors as node attributes.} On both datasets, nodes with higher attribute similarity (i.e., higher AAS with smaller $K$) are more likely to be in the same cluster (i.e., higher RCC). Moreover, above a certain $K$ value, there is no significant difference between the RCC of two random nodes and that of two nodes $v_i$ and $v_j$ such that $v_j$ is the K-nearest neighbor of $v_i$. 
Based on these observations, it is viable to extend KNN augmentation in Section \ref{sec:aug-graph} to an attributed network $\mathcal{N}$ with $n$ nodes and attribute matrix $\XM\in\mathbb{R}^{n\times d}$, by building a KNN augmentation graph $\mathcal{G}_K$ via Eq. \eqref{AK}.

Then we obtain an augmented network $\mathcal{N}_A$ with  topology $\mathcal{N}_O$ and  KNN graph $\mathcal{G}_K$, where $\mathcal{N}_O$ is $(\mathcal{V}, \mathcal{E})$ when $\mathcal{N}$ is an attributed hypergraph $\mathcal{H}$ or $(\mathcal{V}, \mathcal{E}_G)$ for graph $\mathcal{G}$, and $\mathcal{N}_O$ is $(\mathcal{V}, \mathcal{E}_1,\dots,\mathcal{E}_L)$ when $\mathcal{N}$ is an attributed multiplex graph $\mathcal{G}_M$.} 

\extension{\begin{figure}[!t]
\centering
\scalebox{0.75}{
\hspace{-15mm}
\begin{minipage}{0.49\textwidth}
  \centering
  \scalebox{1}{\input{sim-dist-add1}}
\vspace{-2mm}
  \captionof{figure}{\extension{AAS and RCC on Citeseer-DG}}
  \label{fig:sim-dist-add1}
\end{minipage}%
\hspace{-30mm}
\begin{minipage}{0.49\textwidth}
  \centering
 \scalebox{1}{\input{sim-dist-add2}}
\vspace{-2mm}
  \captionof{figure}{\extension{AAS and RCC on ACM}}  \label{fig:sim-dist-add2}
\end{minipage}
}
\vspace{-4mm}
\end{figure}}

\subsection{\extension{Generalized $(\alpha,\beta,\gamma)$-Random Walk}}\label{sec:generalizedrw}
\vspace{-1mm}

\revision{For the augmented network $\mathcal{N}_A = (\mathcal{N}_O, \mathcal{G}_K)$, define $\PM_N$ and $\PM_K$ as the random walk transition matrices of $\mathcal{N}_O$ and $\mathcal{G}_K$ respectively. The generalized  $(\alpha,\beta,\gamma)$-random walk on $\mathcal{N}_A$ is an RWR process over the augmented network $\mathcal{N}_A$, similar to the case of attributed hypergraphs in \mainalgo. The difference from Definition \ref{def:rw-aug-hg} is that when the random walk navigates to another node, with probability $1-\beta_i$, an out-neighbor is drawn from the distribution of $\PM_N$ instead of incident hyperedges. This generalized random walk can also be characterized by the probability in Eq. \eqref{eq:approx-multi-hop}, with transition matrix $\PM$ given as follows.
\begin{equation}\label{eq:p-mat-ancka}
    \PM = (\IM-\BM)\cdot \PM_N + \BM\cdot\PM_K.
\end{equation}

We now formulate $\PM_N$ for different types of networks, including attributed hypergraphs as one special case.
}

\extension{\stitle{Attributed Hypergraph $\mathcal{H}$}
When $\mathcal{N}_O$ is a hypergraph with hyperedge incidence matrix $\HM$, based on Eq. \eqref{eq:p-mat}, $\PM_N$ is shown below. $\PM_N$ considers the transition probability $\PM_V$ from a node to its incident hyperedges and the transition probability $\PM_E$ from each hyperedge to nodes connected by the hyperedge.
\begin{equation}\label{eq:rw-hg}
        \PM_N = \PM_V \PM_E, \text{ where } \PM_V=\DM^{-1}_V\HM\transpose \text{ and } \PM_E=\DM^{-1}_E\HM.
\end{equation}

\noindent\textbf{Attributed Graph $\mathcal{G}$.}
When $\mathcal{N}_O$ is an undirected graph, we can acquire the  transition matrix $\PM_N$ in Eq. \eqref{eq:rw-ag}. If $\mathcal{N}_O$ is directed, we introduce a reversed edge for each edge and consider bidirectional connections between nodes to get $\AM$, $\DM$, and subsequently $\PM_N$.
\begin{equation} \label{eq:rw-ag}
    \PM_N = \DM^{-1}\AM,
\end{equation}
where $\AM$ is the adjacency matrix and $\DM$ is the degree matrix.

\stitle{Attributed Multiplex Graph $\mathcal{G}_M$} When $\mathcal{N}_O$ is a multiplex graph comprising $L$ layers with the same node set $\mathcal{V}$, the $l$-th layer has its own edge set $\mathcal{E}_l$ representing a unique type of connections. 
The overall goal of the clustering task is to make cluster assignments that capture the collective structure of the multiplex graph, transcending the differences across layers.
To achieve this, intuitively, we treat every layer equally and compute $\PM_N$ as in Eq. \eqref{eq:rw-mg}, while layer weighting is left as future work~\cite{FanseuKamhoua2022GRACEAG}.
Given the degree matrix $\DM_l$ and adjacency matrix $\AM_l$  of every $l$-th layer, we get the layer's random walk transition matrix $\DM_l^{-1}\AM_l$, and then compute $\PM_N$ of the multiplex graph by averaging the layer-specific transition matrices. %
Consequently, from the current node $v$, a random walk has $1/L$ probability of selecting each layer $\mathcal{G}_l$, and then within this chosen layer, the next node to visit is picked uniformly at random from the out-neighbors of $v$ in $\mathcal{G}_l$.
\begin{equation} \label{eq:rw-mg}
    \PM_N = \frac{1}{L} \sum_{l=1}^L \DM_l^{-1}\AM_l,
\end{equation}
where $\DM_l$ and $\AM_l$ are the degree matrix and adjacency matrix of the $l$-th layer.

\subsection{\extendagc Algorithm}

With the random walk transition matrix $\PM$ formulated above for various types of attributed networks $\mathcal{N}$, Eq. \eqref{eq:approx-multi-hop} can be reused to calculate $\SM[i,j]$, the probability of a generalized $(\alpha,\beta,\gamma)$-random walk from $v_i$ stopping at $v_j$ in the end.  
The objective function in Section \ref{sec:objfunc} is naturally extended to \extendagc. 
Consequently, our theoretical analysis in Section \ref{sec:objective} remains valid for \extendagc over attributed networks that can be hypergraphs, graphs, and multiplex graphs.

The pseudo-code of \extendagc is outlined in Algorithm \ref{alg:ancka}.
\revision{At Line 1, it obtains transition matrix $\PM_K$ for attribute KNN augmentation. Then as a framework supporting various attributed networks, \extendagc is a generalization of Algorithms \ref{alg:mainalg}-\ref{alg:init} with transition matrix $\PN$ computed depending on the network type at Line 2.  $\PN$ is then used throughout the algorithm as a part of the generalized $(\alpha, \beta, \gamma)$-random walk. The greedy initialization of clusters in Lines 3-11 resembles the procedure in \bcm with the corresponding $\PN$ for RWR simulation.} Since \extendagc needs to pick $k$ nodes in $\mathcal{N}$ with the largest degrees as tentative cluster centers at Line 3 when $\mathcal{N}$ is an attributed multiplex graph, we rank the nodes by their summed degrees across all layers. 

\revision{
Lines 12-24 describe the main clustering process of \extendagc, which extends the hypergraph-specific  Algorithms \ref{alg:mainalg} and \ref{alg:mhcond} with modifications to support attributed graphs and multiplex graphs. First, in orthogonal iterations, calculating $\ZM^{(t)}$ is dependent on the type of $\mathcal{N}$. Second, the MHC objective for general networks stems from the analysis  in Section \ref{sec:objective}, while the formulation with $\PN$ is slightly different.}
In particular, to get MHC $\phi_t$ without materializing the dense matrix $\SM$ in Eq. \eqref{eq:psi-y} that is expensive to compute, we iteratively obtain $\phi_t$ via the intermediate matrix $\FM^{(\ell)}$ in Eq. \eqref{eq:mhc-comp-ancka} at Line 21.
\begin{equation}\label{eq:mhc-comp-ancka}
\FM^{(\ell)} = (1-\alpha) ((\IM-\BM) \PN \FM^{(\ell-1)} + \BM \PM_K \FM^{(\ell-1)}) + \FM^{(0)},
\end{equation}
where $\PN$ is Eq. \eqref{eq:rw-hg}, \eqref{eq:rw-ag}, or \eqref{eq:rw-mg}, depending on the type of $\mathcal{N}$.
\revision{Finally, \extendagc adopts the early stopping criteria in Line 24 and returns the clusters with the lowest MHC obtained. }

\begin{algorithm}[!t]
\caption{\extendagc{}}\label{alg:ancka}
\KwIn{Attributed network $\mathcal{N}$ with KNN augmented graph $\mathcal{G}_K$, the number of clusters $k$, diagonal matrix $\BM$, constant $\alpha$, error threshold $\epsilon_Q$, the numbers of iterations $T_a$, $\gamma$, $T_i$, an integer $\tau$.}
\KwOut{BCM matrix $\YM$}
$\PM_K\gets\DM_K^{-1} \AM_K$\; 
 Get $\PN$ by Eq. \eqref{eq:rw-hg}, \eqref{eq:rw-ag}, or \eqref{eq:rw-mg}, depending on the type of $\mathcal{N}$\;
 $\mathcal{V}_c \gets $ sorted indices of $k$ nodes in $\mathcal{N}$ with $k$ largest degrees\;
 Initialize $\ZM_0\gets \mathbf{0}^{k\times n}$\;
 \lFor{$j\gets 1$ to $k$}{
 $\ZM_0[j,\mathcal{V}_c[j]] \gets 1$
 }
 Initialize $\boldsymbol{\Pi}_c^{(0)} \gets \alpha \ZM_0$\;
 \For{$t \gets 1, 2, \dots T_i$}{
  $\boldsymbol{\Pi}_c^{(t)} \gets (1-\alpha) \boldsymbol{\Pi}_c^{(t-1)} \PN + \boldsymbol{\Pi}_c^{(0)}$\;
 }
 \For{$v_j\in \mathcal{V}$}{
 $g(v_j) \gets \arg\max_{1\le l\le k} \boldsymbol{\Pi}_c^{(T_i)}[l, j]$ \;
 ${\YM}^{(0)}[j,g(v_j)]\gets 1$\;
 }
 $\YM \gets \YM^{(0)}$, $ \widehat{\YM}^{(0)} \gets h({\YM}^{(0)})$\;
 $\QM^{(0)}\gets \frac{\mathbf{1}}{\sqrt{n}}\cdot \mathbf{1} | \widehat{\YM}^{(0)}$ \;
 \For{$t \gets 1, 2, \cdots, T_a$}{
  $\ZM^{(t)} \gets (\IM-\BM)\PN \cdot\QM^{(t-1)} + \BM\PM_K\cdot \QM^{(t-1)}$\;
  $\QM^{(t)}, \RM^{(t)}\gets \texttt{QR}(\ZM^{(t)})$ \;
  \If{t $mod\ \tau = 0$ }{
   ${\YM}^{(t)} \gets $ \texttt{Discretize}($\QM^{(t)}$) \;
   {
    $\widehat{\YM}^{(t)} \gets h(\YM^{(t)});\ \FM^{(0)} \gets \alpha\widehat{\YM}^{(t)}$\;
 \For{$\ell \gets 1, 2, \dots \gamma$}{
  Compute $\FM^{(\ell)}$ according to Eq. \eqref{eq:mhc-comp-ancka}
 }
 $\Phi({\YM}^{(t)}) \gets 1-\frac{1}{k}trace(\widehat{\YM}^{(t)\top} \FM^{(\gamma)})$\;
   \lIf{$\Phi({\YM}^{(t)})<\Phi({\YM})$}{$\YM \gets \YM^{(t)}$}
   \lIf{Eq. \eqref{eq:tcond-1} or Eq. \eqref{eq:tcond-2} holds}{\textbf{break}} 
   }
  }
 }
 {
 \Return ${\YM}$\;
 }
\end{algorithm}

\stitle{Complexity} %
When $\mathcal{N}$ is an attributed graph, constructing transition matrix $\PN$ takes $O(n\overline{\delta})$ time, where $\overline{\delta}$ is the average node degree. 
For a multiplex network $\mathcal{N}$ with $L$ layers, the previous results are still valid when $L$ is regarded as constant, as $\PN$ is aggregated from the transition matrices of all simple graph layers.
Given that the number of nonzero entries in $\PN$ is subject to $O(n\overline{\delta})$, \extendagc (Algorithm \ref{alg:ancka}) has the same complexity as Algorithm \ref{alg:mainalg}. According to our analysis in Section \ref{sec:analysis}, the time complexity of \extendagc is $O(kn(\overline{\delta}+K+k))$ while its space complexity is $O(n(\overline{\delta}+K+k))$. Since $k$ and $K$ can be viewed as constants, \extendagc has space and time complexity of $O(n\overline{\delta})$.

\section{GPU-Accelerated \extendgpu}\label{sec:extendgppu}
On large attributed networks, e.g., Amazon and MAG-PM hypergraphs,  each with more than 2 million nodes, as reported in Table \ref{tab:efficiency-result}, \mainalgo with 16 CPU threads still needs 1286s and 1372s respectively for clustering, despite its superior efficiency compared with baselines. 
Moreover, \mainalgo does not exhibit acceleration proportional to increased CPU threads. 
As shown in Figure \ref{fig:multi}, when the number of CPU threads is raised from 1 to 32, the time drops from around 3000s to 1200s, with a speedup of merely 2.5 (Amazon) or 2.7 (MAG-PM). 
In particular, increasing the number of threads from 16 to 32 provides rather limited acceleration (less than 10\%).

\revision{To overcome the limitation of CPU parallelization, we resort to the massive parallel processing power of GPUs  (graphical processing unit) and develop \extendgpu to boost efficiency, with about one order of magnitude speedup on large networks with millions of nodes in experiments. For example, \extendgpu only needs 120s on an MAG-PM dataset, over 10 times faster than the 1372s of \extendagc. Compared to CPUs, the design of GPUs enables them to leverage numerous threads to handle data processing simultaneously, which is beneficial for vector and matrix operations at scale. Please see \cite{cook2012cuda} for details on GPU computing.
}

As shown in Figure \ref{fig:runtime-vs} of  Section \ref{sec:runtime-als} for runtime analysis, the major time-consuming components of \extendagc include invoking \discr (Line 18 in Algorithm \ref{alg:ancka}), the construction of KNN graph $\mathcal{G}_K$, and expensive matrix operations in orthogonal iterations, greedy initialization and MHC evaluation.
With the CuPy library, matrix operations throughout Algorithm \ref{alg:ancka} can be done on GPUs more efficiently.
In the following, we elaborate on the GPU-based discretization and $\mathcal{G}_K$ construction techniques adopted in \extendgpu.

\stitle{GPU-based Discretization \discrgpu} \extendagc uses the off-the-shelf \discr approach~\cite{yu_multiclass_2003} to compute discrete cluster labels $\YM$ from real-valued eigenvectors $\QM$, which could cost substantial time on large datasets. 
Here, we develop a CUDA kernel \discrgpu for efficiency. 
In what follows, we first explain how the discretization algorithm improves the optimization objective in Definition \ref{def:discr}, and then present the design of \discrgpu in Algorithm \ref{alg:discr}.

Given an eigenvector matrix $\QM$ with its row-normalized matrix ${\tilde{\QM}}$, discretization is aimed to find a discrete solution $\YM_{opt}$ that minimizes the objective in Definition \ref{def:discr}.

\begin{definition}\label{def:discr} (Discretization~\cite{yu_multiclass_2003}) The solution to the following optimization problem is the optimal discrete $\YM_{opt}$. %
\begin{gather*}
\YM_{opt}=\argmin_\YM ||\YM-{\tilde{\QM}}\RM||_F^{2} \\
\text{\it s.t. } \YM \in \{0,1\}^{n\times k},\ \YM \mathbf{1}_k =\mathbf{1}_n,\ \RM \in \mathbb{R}^{k\times k},\ \RM^{\TM} \RM=\IM_k,
\end{gather*}
where  ${\tilde{\QM}}$  is the row-normalized matrix of an eigenvector matrix $\QM$, $\RM$ is a rotation matrix, and $||\mathbf{M}||_F$ denotes the Frobenius norm of matrix $\mathbf{M}$.

\end{definition}

The \discr approach finds a nearly global optimal solution by alternately updating one of $\YM$ and $\RM$ while keeping the other fixed. With $\RM$ fixed, $\YM[i,l]$ is updated to 
\begin{equation}\label{eq:updateY}
    \YM[i,g]=\begin{cases}
1, &\text{ if } g = \arg\max_{1\leq j \leq k} (\tilde{\QM} \RM)[i, j] \\
0, & \text{otherwise.}
\end{cases}
\end{equation}

With $\YM$ fixed, $\tilde{\YM}$ is the column-normalized matrix of $\YM$, and $\RM$ can be updated as follows with SVD decomposition. 
\begin{equation}\label{eq:svd-in-discr}
\RM=\VM\UM\transpose\text{, where }  \UM\OmegaM \VM\transpose \text{ is an SVD of } \tilde{\YM}\transpose \tilde{\QM}. 
\end{equation}

The iterative process can terminate early when an objective value $obj$ based on $\OmegaM$ converges, i.e., its change over the last iteration is within machine precision. This objective is calculated as $obj=n-2\times trace(\OmegaM)$~\cite{yu_multiclass_2003}.

\begin{figure}[!t]
    \extension{
    \centering
    \resizebox{0.52\columnwidth}{!}{
        \input{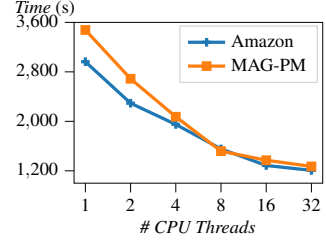}
    }
    \vspace{-2mm}
    \caption{\extension{Runtime of \mainalgo with CPU parallelization}}
    \label{fig:multi}}
    \vspace{-4mm}
\end{figure}

\begin{algorithm}[!t]
\caption{\discrgpu}\label{alg:discr}
\KwIn{eigenvector matrix $\QM$}
\KwOut{Intermediate BCM matrix $\YM$}
 \textbf{Parallel} \For{$i\gets 1,2,\cdots, n$}{ $ \tilde{\QM}[i] \gets \frac{\QM[i]}{\Vert \QM[i] \Vert_{2}}$ \;}
 $\RM \gets \IM_{k}$ \; 
 \While{$iter\gets 1,2,\cdots,max\_iter$}{ 
    Update $\YM$ by Eq. \eqref{eq:updateY} via argmax kernel on GPU\;
    \textbf{Parallel} \For{$j\gets 1,2,\cdots, k$}{$col\_sum[j]\gets\sum_{i=1}^n \YM[i,j]$}
    \textbf{Parallel} \For{each $tid<k$ in blocks}{ 
   $\tilde{\YM}[bid,tid] \gets \frac {\YM[bid,tid]} {col\_sum[tid]}$\; }
   $\UM,\OmegaM,\VM\transpose \gets \texttt{SVD\_GPU}(\tilde{\YM}\transpose \tilde{\QM})$ %
   \;
    $\RM \gets \VM\UM\transpose$ on GPU\;
   \lIf{\text{Objective value} $obj$ \text{does not change}}{\textbf{break}}
   }
   {
  \Return  $\YM$\;
  }

\end{algorithm}

We implement the CUDA kernel \discrgpu in Algorithm \ref{alg:discr} to perform the process above to obtain $\YM$. In details,
\discrgpu leverages the grid-block-thread hierarchy of GPU to assign threads to handle $n\times k$ matrices, including $\QM$ and $\YM$. 
Each row in such a matrix is processed by a block of threads, identified by a block id $bid$; each of the $k$ elements in the row is handled by a thread $tid$ in the block. Consequently, given a matrix $\QM$,  we can use $\QM[bid,tid]$ to represent that the corresponding element in $\QM$ is handled by the $tid$-th thread in block $bid$ on a GPU.
Parallel row normalization is performed at Lines 1-2 to get $\tilde{\QM}$. 
After initializing $\RM$ as a $k\times k$ identity matrix (Line 3), we alternately update $\tilde{\YM}$ and $\RM$ for at most $max\_iter$ iterations (Lines 4-12) and terminate early when the objective value $obj$ does not change over the current iteration at Line 12.
Within an iteration, we first update $\YM$ at Line 5, then perform column normalization to get $\tilde{\YM}$ (Lines 6-9), and then perform SVD on GPU over $\tilde{\YM}\transpose \tilde{\QM}$ to get $\UM$ and $\VM$ at Line 10, which helps to update $\RM$ at Line 11.
Finally, $\YM$ is returned at Line 13.

\stitle{KNN construction} 
An $n\times d$ attributed matrix $\XM$ requires KNN search on its rows to construct the augmented graph $\mathcal{G}_K$ and thus the transition matrix $\PM_K$.
For this purpose, we adopt Faiss \cite{johnson2019billion}, a GPU-compatible similarity search library. 
In Algorithm \ref{alg:knn} for $\mathcal{G}_K$ construction, we first normalize all rows in $\XM$ at Lines 1-2 to facilitate the computation of cosine similarity between row vectors. 
Faiss supports various indexes for KNN computation, and the index type suitable for \extendagc is determined based on the input data volume. 
For small or medium datasets where the number of nodes $|\mathcal{V}|$ is below 100,000, since the time cost for exact similarity search is affordable, we choose the flat index with a plain encoding of each row vector in $\XM$, to achieve exact KNN computation (Lines 3-4). Otherwise, we turn to approximate nearest neighbor search on large datasets with the IVFPQ index that combines the inverted file index (IVF) with the product quantization (PQ) technique at Line 6. In particular, IVF index narrows down the search to closely relevant partitions that contain the nearest neighbors at a high probability, while PQ produces memory-efficient encoding of attribute vectors. %
Faiss on GPU is invoked to get the KNN of each row in $\XM$, and $\AM_K$ is obtained by Eq. \eqref{AK} at Lines 7-8. 
Then, the degree matrix $\DM_K$ and transition matrix $\PM_K$ are computed on GPU (Lines 9-10) and returned at Line 11.
}

\begin{algorithm}[t]
\caption{GPU-based  $\mathcal{G}_K$ construction}\label{alg:knn}
\KwIn{Network $\mathcal{N}$, attribute matrix $\XM$, parameter $K$.}
\KwOut{KNN transition matrix $\PM_K$}
\textbf{Parallel} \For{$i\gets 1,2,\cdots, n$}  {$\XM[i] \gets \frac{\XM[i]}{||\XM[i]||_2}$\;}  
 \If{$|\mathcal{V}|<100,000$}{$index \gets$ \texttt{FlatIndex} $(\XM)$ \;}
 \Else {$index \gets$ \texttt{IVFPQIndex} $(\XM)$ \;}
 Invoke Faiss on GPU to get the KNN of each row in $\XM$ \;
 Get $\AM_{K}$ by Eq. \eqref{AK} on GPU \;
 $\DM_{K} \gets \texttt{Diag}(\AM_{K} \mathbf{1}_n)$ on GPU \;
 $\PM_{K} \gets \DM_{K}^{-1}\AM_{K}$  on GPU\;
 \Return  $\PM_{K}$\;
\end{algorithm}

%% file: experiment.tex
\section{Experiments}\label{sec:experiment}
\extension{We evaluate the proposed \extendagc and competitors
in terms of clustering quality and efficiency. We also evaluate the performance of \extendgpu on all clustering tasks.  
In experiments, we uniformly refer to our method as \extendagc while making it clear in the context whether \extendagc is for AHC (i.e., \mainalgo), AGC, or AMGC.}
All the experiments are conducted on a Linux machine powered by Intel Xeon(R) Gold 6226R CPUs, 384GB RAM, and NVIDIA RTX 3090 GPU. A maximum of 16 CPU threads are available if not otherwise stated. The code is at \url{https://github.com/gongyguo/ANCKA}.

\subsection{Experimental Setup}
\subsubsection{Datasets}
\input{dataset}

Table \ref{tab:dataset} provides the statistics of \extension{17 real-world attributed networks used in experiments, including attributed hypergraphs (HG), undirected graphs (UG), directed graphs (DG), and multiplex graphs (MG). $|\mathcal{V}|$ and $|\mathcal{E}|$ are the number of nodes and edges (or hyperedges), respectively, $d$ is the attribute dimension and $k$ is the number of ground-truth clusters.}

We gather 8 attributed hypergraph datasets. Query dataset \cite{whangMEGAMultiviewSemisupervised2020} is a   Web query hypergraph, where nodes represent queries and are connected by hyperedges representing query sessions, \revision{and nodes are associated with attributes of keyword embeddings and associated webpages.} Cora-CA, Cora-CC, Citeseer, and DBLP are four benchmark datasets used in prior work~\cite{yadati2019hypergcn}. All of them are originally collected from academic databases, where each node represents a publication, node attributes are binary word vectors of abstract, and research topics are regarded as ground-truth clusters. Hyperedges correspond to co-authorship in Cora-CA and DBLP datasets or co-citation relationship in Cora-CC and Citeseer datasets. 20News dataset \cite{heinTotalVariationHypergraphs2013} consists of messages taken from Usenet newsgroups. Messages are nodes, and the messages containing the same keyword are connected by a corresponding hyperedge, and the TF-IDF vector for each message is used as the node attribute.
Amazon dataset is constructed based on the 5-core subset of Amazon reviews dataset~\cite{ni2019justifying}, where each node represents a product and a hyperedge contains the products reviewed by a user. For each product, we use the associated textual metadata as the node attributes and the product category as its cluster label.
MAG-PM dataset is extracted from the Microsoft Academic Graph~\cite{sinhaOverviewMicrosoftAcademic2015a}, where nodes, co-authorship hyperedges, attributes, and cluster labels are obtained as in other academic datasets (i.e., Cora-CA, Cora-CC, Citeseer, and DBLP).

\extension{In Table \ref{tab:dataset}, we also consider 6 attributed graphs, which are commonly used for AGC \cite{zhangAGC10.5555/3367471.3367643, FanseuKamhoua2022GRACEAG, yangEffectiveScalableClustering2021, Chiang2019ClusterGCNAE}.
Cora, Citeseer-UG, Wiki, and Amazon2M are undirected, while Citeseer-DG and TWeibo are directed. 
TWeibo \cite{yangEffectiveScalableClustering2021} and Amazon2M \cite{Chiang2019ClusterGCNAE} are two large-scale attributed graphs. TWeibo is a social network where each node represents a user, and the directed edges represent relationships between users. Amazon2M is constructed based on the co-purchasing networks of products on Amazon.
Cora, Citeseer-UG, and Citeseer-DG are citation networks where nodes represent publications, a pair of nodes are connected if one cites the other, and nodes are associated with binary word vectors as features. Wiki is a webpage network where each edge in the graph indicates that one webpage is linked to the other, while the node attributes are TF-IDF feature vectors. 
Moreover, three attributed multiplex graphs, namely ACM, IMDB, and DBLP-MG, are considered for AMGC \cite{Jing2021HDMIHD, Pan2021MultiviewCG, Lin2021MultiViewAG}.  
\revision{ACM is an academic publication network comprising co-author  and co-subject  graph layers, as well as bag-of-words attributes of keywords. IMDB is a movie network with plot text embeddings as attributes and two graph layers representing the co-director (directed by the same director) and co-actor  (starring the same actor) relations, respectively. DBLP-MG is a researcher network including publication keyword vectors as attributes and three graph layers: co-author, co-conference
(publishing at the same conference),
and co-term (sharing common key terms). ACM and DBLP-MG have research areas labeled as ground truth clusters, while IMDB is labeled by movie genres. }
}

\subsubsection{Competitors and Parameter Settings}\label{sec:mainalg-params}
\vspace{-1mm}
{The \revision{19} competitors for AHC are summarized as follows: 
\vspace{-\topsep}
\begin{itemize}[leftmargin=*]
\item 3 plain hypergraph clustering methods including \hncut \cite{zhouLearningHypergraphsClustering2007}, \hyperadj \cite{rodriguez_laplacian_2002}, and \kahypar~\cite{MT-KaHyPar-Q-F}; 
\item the extended AHC versions of the 3 methods above (dubbed as \athncut, \athyperadj, and \atkahypar), which work on an  augmented hypergraph with attribute-KNN hyperedges of all nodes merged into the input hypergraph;
\item  \atmetis that applies the traditional graph clustering algorithm \texttt{Metis} \cite{karypisMultilevelkwayPartitioningScheme1998} over a graph constructed by clique expansion of the input hypergraph and attribute KNN graph augmentation; \revision{\infomap~\cite{rosvallMapEquation2009}, \louvain~\cite{blondel2008fast}, \kmqi and \knibble (extended from \texttt{MQI} \citep{lang2004flow} and {PageRank-Nibble} \citep{andersen2006local} for $k$-way clustering via $k-1$ consecutive bisections as described in technical report \cite{report}) on the same KNN-augmented clique-expansion graph;} 
\item 3 AHC algorithms including the recent \grac~\cite{fanseukamhouaHyperGraphConvolutionBased2021} and NMF-based approaches (\gnmfl~\cite{cai2010graph, fanseukamhouaHyperGraphConvolutionBased2021} and \jnmf~\cite{duHybridClusteringBased2019});
\item \arwc and \arws, obtained by applying an attributed graph clustering method \arw~\cite{yangEffectiveScalableClustering2021} over the graphs reduced from hypergraphs by clique expansion and star expansion, respectively;  \revision{probabilistic model \cesna~\cite{yang2013community} with clique-expansion;}
\item \revision{\kmeans and \aggcluster (hierarchical agglomerative clustering \cite{wardjr.HierarchicalGroupingOptimize1963}) algorithms applied to the node attribute matrix.}
\end{itemize}}

\extension{To evaluate the \extendagc framework, we compare \revision{16} competitors for AGC, including \kmeans, \revision{\aggcluster} and the following: 
\vspace{-\topsep}
\begin{itemize}
\item \revision{6} AGC approaches including NMF-based algorithm \gnmf \cite{cai2010graph}, graph convolution algorithm \agc \cite{zhangAGC10.5555/3367471.3367643} , \revision{probabilistic model \cesna \cite{yang2013community}}, spectral clustering on fine-grained graphs method \fgc \cite{Kang2022FinegrainedAG}, attributed random walk approach \arw \cite{yangEffectiveScalableClustering2021}, and the clustering framework \grace \cite{FanseuKamhoua2022GRACEAG} generalized from \grac.
    \item \ncut\cite{shi2000normalizedcuttpami} and \metis\cite{karypisMultilevelkwayPartitioningScheme1998} that are conventional graph clustering methods applied to the input graph;
    \item \atncut and \atmetis   that are \ncut and \metis applied to
    the augmented graph with attribute KNN; \revision{\infomap \cite{rosvallMapEquation2009}, \louvain \cite{blondel2008fast}, \kmqi \citep{lang2004flow} and \knibble \citep{andersen2006local} on the augmented graph with attribute KNN.}    
\end{itemize}}

\extension{
We compare \extendagc with \revision{16} competitors for AMGC task, including \kmeans, \revision{\aggcluster} and the following: 
\vspace{-\topsep}
\begin{itemize}
 \item 5 AMGC methods: a multi-view graph auto-encoder model \omac \cite{Fan2020One2MultiGA}, \hdmi \cite{Jing2021HDMIHD} that learns node embeddings via higher-order mutual information loss, \mcgc \cite{Pan2021MultiviewCG} and \magc \cite{Lin2021MultiViewAG} which perform graph filtering and find a consensus graph for spectral clustering, and \grace \cite{FanseuKamhoua2022GRACEAG} that is a general graph convolution clustering method; 
    \item \ncut\cite{shi2000normalizedcuttpami} and \metis\cite{karypisMultilevelkwayPartitioningScheme1998} that apply traditional graph clustering methods over the aggregation of the adjacency matrices of all graph layers in the input multiplex graph;
    \item \atncut and \atmetis that apply \ncut and \metis to the aggregated matrix of all layers' adjacency matrices and the attribute KNN graph; \revision{\infomap \cite{rosvallMapEquation2009}, \louvain \cite{blondel2008fast}, \kmqi \citep{lang2004flow} and \knibble \citep{andersen2006local} in the same way;}  
    \item \revision{\cesna \cite{yang2013community} that treats the aggregated adjacency matrix of all layers as an attributed graph;}
   
\end{itemize}}

\input{results-revision.tex}

For all competitors, we adopt the default parameter settings as suggested in their respective papers. \extension{Hyperparameters for AMGC algorithms \mcgc and \magc are tuned as instructed in the corresponding papers, and we report the best results acquired.}
 As for \extendagc on attributed hypergraphs, i.e., \mainalgo \cite{Li2023EfficientAE}, unless otherwise specified, we set parameters on all datasets: $\alpha=0.2$, $\beta=0.5$, and $\gamma=3$, parameter $K=10$ for KNN construction, the convergence threshold $\epsilon_Q = 0.005$, and the numbers of iterations $T_a=1000$, $T_i=25$.
The interval parameter $\tau$ is set to $5$ on all datasets except the large and dense hypergraph Amazon, where we set $\tau=1$ to expedite early termination in light of the immense per-iteration overhead when processing Amazon. On large datasets (i.e., Amazon and MAG-PM), $T_i$ is set to $1$ and $\beta=0.4$. 
\extension{In \extendagc, for attributed graphs and multiplex graphs, we fix $K=50$, except for large datasets TWeibo and Amazon2M with $K=10$. \revision{In particular, we find it necessary to adjust the $\beta$ parameter 
for certain instances following the practice in recent works  \citep{FanseuKamhoua2022GRACEAG,Pan2021MultiviewCG,Lin2021MultiViewAG}. $\beta$ is set to $0.5$ for Cora and Wiki and $0.4$ on Citeseer-UG, Citeseer-DG, TWeibo, and Amazon2M. We tune $\beta$ in $[0.1, 0.9]$ by step size 0.1 for multiplex graphs.
}
All the remaining hyperparameters in \extendagc follow the default setting of \mainalgo. The parameter settings in GPU-based \extendgpu are identical to \extendagc}.

\input{performance1}

\begin{table*}[t]
  \caption{\extension{Attributed Multiplex Graph Clustering (AMGC) Quality.}}
  \label{tab:performance1.3}
  \centering
  \vspace{-2mm}
  \extension{
  \small
  \resizebox{0.96\textwidth}{!}{
   \setlength{\tabcolsep}{7pt}
   \renewcommand{\arraystretch}{0.93} %

  \begin{tabular}{=c|+c+c+c+c|+c+c+c+c|+c+c+c+c|c}
    \toprule
    & \multicolumn{4}{c|}{{{{ACM}}}}& \multicolumn{4}{c|}{{{IMDB}}} & \multicolumn{4}{c|}{{{DBLP-MG}}} &{\revision{\textbf{Quality}}} \\
    {Algorithm} & Acc            & F1             & NMI            & ARI   & Acc            & F1             & NMI            & ARI    & Acc            & F1             & NMI            & ARI  &{\revision{\textbf{Rank}}} \\ \midrule
    {\metis} & 0.648 & 0.651 & 0.389  & 0.369  & 0.376 & 0.374 & 0.004& 0.004   & 0.864 & 0.860& 0.660 & 0.688 &11.6 \\
    {\ncut} & 0.350 & 0.174 & 0.003 & 0.000 & 0.378 & 0.185 & 0.002 & 0.000& 0.299 & 0.125 & 0.012 & -0.001 &15.3\\
    {\kmeans} & 0.679 & 0.681 & 0.320 & 0.312 & 0.525 & 0.531& 0.146 &  0.139 & 0.368  & 0.285 & 0.083 & 0.060 &10.1\\
    \rowstyle{\color{\revcolor}}
    \aggcluster &0.576 &0.557 &0.234 &0.222  &0.483 &0.462 &0.100 & 0.101 & 0.381 &0.304 &0.131 &0.070 &11.4\\
    \midrule
    {\atmetis} & 0.755 & 0.757 & 0.510  & 0.490 & 0.546 & \underline{0.551} & 0.161 & 0.152 & 0.868 & 0.864 & 0.669 & 0.697 &6.5 \\
    {\atncut}& 0.778 & 0.775 & 0.462  & 0.465  & 0.499 & 0.466 & 0.154 & 0.165  & 0.360 & 0.285 & 0.104 & 0.023 &8.8 \\
    \rowstyle{\color{\revcolor}}
    \kmqi &0.351 &0.174& 0.001 &0.000 &0.377 &0.183 &0.001&0.000 &0.295 &0.115 &0.002 &0.000 &15.8\\
    \rowstyle{\color{\revcolor}}
    \knibble &0.343 &0.221 &0.018 &0.001 &0.370 &0.251& 0.022& 0.005 &0.295 &0.115 &0.002 &0.000 &15.2\\ 
    \rowstyle{\color{\revcolor}}
    \infomap   &0.653 &0.665 &0.418 &0.353 &0.412 &0.362 &0.027 &0.025 &0.296 &0.116 &0.002 &0.000 &12.6\\ 
    \rowstyle{\color{\revcolor}}
    \louvain  &0.659 &0.670 &0.422 &0.364  &0.452 &0.392 &0.057 &0.065 & 0.909& 0.900 &0.731 &0.788 &8.1 \\
    \rowstyle{\color{\revcolor}}
     {\cesna} &0.624 &0.593 &0.405 &0.330 &0.377 &0.329 &0.006 &0.007  &0.827 &0.820 &0.583 &0.603 &12.0\\
    \midrule
    {\omac} & 0.895 & 0.897 & 0.667  & 0.716 & 0.547 & {0.550} & 0.135 & 0.139  & 0.873 & 0.865 & 0.669 & 0.705 &5.5  \\
    {\hdmi} & 0.900 & 0.899 & 0.695  & 0.732  & 0.541 & 0.547 & 0.162 & 0.142 & 0.895 & 0.885 & 0.706 & 0.761 &4.7\\
    {\mcgc} & \underline{0.915} & \underline{0.916} & \underline{0.709}  & \underline{0.763} & {0.567} & 0.545 & 0.164 & {0.186}  & 0.902 & 0.895 & 0.716 & 0.771 &3.5\\
    {\magc} & 0.872 & 0.872 & 0.597  & 0.659  & 0.484 & 0.424 & 0.057 & 0.062  & \underline{0.928} & \underline{0.923} & \underline{0.771} & \underline{0.827} &6.0\\
    {\grace} & 0.889 & 0.891 & 0.651  & 0.698 & \textbf{0.629} & \textbf{0.629} & \textbf{0.185} & \textbf{0.205} & 0.923 & 0.918 & 0.767 & 0.817 &\underline{3.0} \\
    {\extendagc} & \textbf{0.928} & \textbf{0.928} & \textbf{0.739}  & \textbf{0.796}  & \underline{0.576} & 0.544 & \underline{0.176} & \underline{0.195}  & \textbf{0.933} & \textbf{0.929} & \textbf{0.785} & \textbf{0.839} &\textbf{1.7}\\ 
    \bottomrule
  \end{tabular}}}
\end{table*}

\subsection{Performance Evaluation}
In this section, we report clustering quality and efficiency of all methods on all datasets. For each method, we repeat 10 times and report the average performance. 

\subsubsection{Quality Evaluation}
\vspace{-1mm}

The clustering quality is measured by 4 classic metrics including overall accuracy (Acc), average per-class F1 score (F1), normalized mutual information (NMI), and adjusted Rand index (ARI). The former three metrics are in the range $[0,1]$, {whereas ARI ranges from -0.5 to 1.} \revision{We also sort all methods by each metric and calculate their average Quality Rank for AHC, AGC, and AMGC, provided in the last column of Tables \ref{tab:large-result}, \ref{tab:performance1.2} and \ref{tab:performance1.3}.}

\input{results-efficiency-revision.tex}

\stitle{AHC}
Tables \ref{tab:small-result} and \ref{tab:large-result} present the Acc, F1, NMI, and ARI scores of each method on small and medium/large {attributed hypergraph} datasets, respectively. 
{The first observation from Tables \ref{tab:small-result} and \ref{tab:large-result} is that \extendagc on attributed hyergraphs (i.e., \mainalgo) consistently achieves outstanding performance over all competitors on all datasets under almost all metrics, often by a significant margin. 
\revision{\extendagc has a quality rank of {1.3}, much higher than the runner-up \atmetis (4.9) and \grac (5.2).}
On all the four small datasets (\textit{i.e.}, Query, Cora-CA, Cora-CC, and Citeseer),  \revision{\extendagc outperforms the best competitors (underlined in Table \ref{tab:small-result}) by at least 1.9\% in terms of Acc and NMI.}
On all the four medium/large attributed hypergraphs (\textit{i.e.}, 20News, DBLP, Amazon, and MAG-PM), \extendagc also yields remarkable improvements upon the competitors, with percentages up to {12.6\%}, 10.4\%, 6.5\%, {13.6\%} in Acc, F1, NMI, and ARI respectively. 
Few exceptions exist, where \extendagc still leads in three out of the four metrics, demonstrating the best overall performance.
The results in Tables \ref{tab:small-result} and \ref{tab:large-result} also confirm the effectiveness of \extendagc over various attributed hypergraphs from different application domains, \textit{e.g.}, web queries, news messages, and review data. } The {performance} of \extendagc is ascribed to our  optimizations based on KNN augmentation and MHC in Section \ref{sec:AuHC} and Section  \ref{sec:objective}, and the framework for generating high-quality BCM matrices in Section \ref{sec:algo}.

\extension{\stitle{AGC} Tables \ref{tab:performance1.1} and \ref{tab:performance1.2} present the Acc, F1, NMI, and ARI scores of each method on all attributed graphs for AGC task. 
\extendagc consistently outperforms existing competitors under most metrics, though few exceptions exist where \extendagc is comparable to the best. 
\revision{\extendagc has a quality rank of {1.3}, much higher than the runner-up with quality rank 4.0.} For example, on Citeseer-UG in Table \ref{tab:performance1.1}, \extendagc achieves higher Acc, F1, NMI and ARI than the runner-up performance underlined. 
On the two large datasets, TWeibo and Amazon2M in Table \ref{tab:performance1.2}, \extendagc also produces clusters with high quality, while \gnmf, \agc, and \fgc run out of memory or cannot finish within 12 hours.
Notably, on Amazon2M, \extendagc surpasses all methods on all metrics except F1 (0.006 behind \atmetis) while achieving 0.494 accuracy (\revision{runner-up is \louvain at 0.463}) and 0.545 ARI (\revision{runner-up is \louvain at 0.520}).
The effectiveness of \extendagc validates the versatility of the proposed techniques for different clustering tasks, e.g., AGC.
Besides, \atmetis and \atncut generally outperform \metis and \ncut in AGC performance, respectively, exhibiting the efficacy of the proposed KNN augmentation.}

\extension{\stitle{AMGC} Table \ref{tab:performance1.3} reports the Acc, F1, NMI, and ARI scores of all methods on all attributed multiplex graphs. \revision{\extendagc has the best quality rank.} As shown, on ACM and DBLP-MG, \extendagc achieves the best clustering quality among all methods under all metrics, with NMI and ARI leading by at least 3\% on ACM, 
while being the second best in three metrics on IMDB. As shown later in Table \ref{tab:performance2.1}, on these datasets, \extendagc is faster than existing native AMGC methods by at least an order of magnitude.
With the intuitive design of random walk transition matrix $\PN$ on multiplex graphs in Section \ref{sec:generalizedrw}, \extendagc can utilize the proposed KNN augmentation, clustering objective, and optimization techniques to maintain its excellent performance on the AMGC task.}

\input{performance2.tex}

\begin{table}[!t]
\centering
\caption{\extension{Efficiency of Attributed Multiplex Graph Clustering (AMGC) Algorithms (Time in Seconds, RAM in GBs). The Quality Rank column is from Table \ref{tab:performance1.3}. Among all native AMGC methods  in the last 6 rows, the best is in bold, and the runner-up is underlined.}}
\label{tab:performance2.1}
\vspace{-2mm}
\small
\extension{
\resizebox{1\columnwidth}{!}{
\setlength{\tabcolsep}{4pt}
\renewcommand{\arraystretch}{0.96} %

\begin{tabular}{=c|+c+c|+c+c|+c+c|c}
\toprule
  & \multicolumn{2}{c|}{{{ACM}}} & \multicolumn{2}{c|}{{IMDB}} & \multicolumn{2}{c|}{{DBLP-MG}} & Quality \\
{Algorithm}          & Time           & RAM   & Time          & RAM   & Time         & RAM  &Rank   \\ \midrule
\rowstyle{\color{\revcolor}} \metis &0.477 &0.382 &0.037 &0.375 &1.798 &0.602 &11.6 \\
\rowstyle{\color{\revcolor}} \ncut &0.761 &0.392 &0.123 &0.384 &2.218 &0.611 &15.3\\
\rowstyle{\color{\revcolor}} \atmetis &1.418 &1.034 &1.181 &1.134 &2.441 &0.672 &6.5 \\
\rowstyle{\color{\revcolor}} \atncut &1.324 &1.037 &1.236 &1.141 &2.587 &0.675 &8.8\\ 
\rowstyle{\color{\revcolor}} \kmqi &1.033 &1.143 &1.064 &1.319 &1.048 &0.957 &15.8\\
\rowstyle{\color{\revcolor}} \knibble &7.230 &0.696 &10.32 &0.766 &3.999 &1.109 &15.2\\
\rowstyle{\color{\revcolor}} \infomap &17.78 &1.547 &3.624 &1.260 &48.45 &3.883  &12.6\\
\rowstyle{\color{\revcolor}} \louvain &43.91 &1.300 &9.537 &1.151 &158.0 &3.948 &8.1\\
\rowstyle{\color{\revcolor}} \cesna  &68.85 &0.309 &32.28 &0.372 &819.2 &0.534 &12.0\\
\midrule
        \omac &115.0 &1.691 &679.1 &2.109 &684.1 &2.638 &5.5\\
        \hdmi &161.2 &2.902 &245.9 &2.980 &537.8 &3.162&4.7 \\
        \mcgc &748.2 &1.697 &1552 &2.414 &2245 &3.283 &3.5\\
        \magc &\underline{26.10} &1.301 &33.69 &1.908 &\underline{35.98} &2.665 &6.0 \\
        \grace &110.1 &\underline{1.173} &\underline{21.81} &\textbf{1.341}&49.33 &\textbf{0.672} &\underline{3.0} \\
        \extendagc &\textbf{1.738} &\textbf{1.062} &\textbf{1.574} &\underline{1.485} &\textbf{3.766} &\underline{0.691} &\textbf{1.7}\\
 \bottomrule
\end{tabular}}}
\vspace{-2mm}
\end{table}

\subsubsection{Efficiency Evaluation}
\revision{Tables \ref{tab:efficiency-result}, \ref{tab:performance2} and \ref{tab:performance2.1} report the runtime (in seconds, with KNN construction included) and memory overhead (in Gigabytes),  for AHC, AGC, and AMGC, respectively.
For ease of comparing the trade-off between quality and efficiency, the last column of Tables \ref{tab:efficiency-result}, \ref{tab:performance2} and \ref{tab:performance2.1} contains the corresponding quality ranks from Tables \ref{tab:large-result}, \ref{tab:performance1.2} and \ref{tab:performance1.3}, respectively.
In each table, the methods are separated into two categories: \textit{non-native} methods extended from other clustering problems and \textit{native} methods for the corresponding task. For instance, in Table \ref{tab:efficiency-result}, there are 4 native AHC methods in the last 4 rows, while the non-native methods are in the rows above. 

In Tables \ref{tab:efficiency-result}, \ref{tab:performance2}, and \ref{tab:performance2.1}, although certain non-native methods are efficient, their quality ranks in terms of clustering quality are typically low. 
Hence, in the following, we mainly compare the efficiency of \extendagc against the native methods for each task. A method is terminated early if it runs out of memory (OOM) or cannot finish within 12 hours. 

}

\noindent\textbf{AHC.} 
In Table \ref{tab:efficiency-result}, compared with native AHC methods, we can observe that \extendagc is significantly faster on most datasets, often by orders of magnitude. 
For example, on a small graph Citeseer, \extendagc takes $0.635$ seconds, while the fastest AHC competitor \grac needs $13.15$ seconds, meaning that \extendagc is $20.7\times$ faster. On large attributed hypergraphs including Amazon and MAG-PM, most existing AHC solutions fail to finish due to the OOM errors, whereas \extendagc achieves $11.4\times$ and $2.6\times$ speedup over the only viable native AHC competitor \grac on Amazon and MAG-PM, respectively. { An exception is 20News, which contains a paucity of hyperedges (100 hyperedges), where \extendagc is slower than \grac. Recall that in Table \ref{tab:large-result}, compared to \extendagc, \grac yields far inferior accuracy in terms of clustering on 20News, which \revision{highlights the advantages}  of \extendagc over \grac.} \revision{
Additionally, while \atmetis is fast, it achieves an average quality rank of 4.9, which falls short of the 1.7 quality rank attained by \extendagc. As shown in Tables \ref{tab:small-result} and \ref{tab:large-result}, \extendagc surpasses \atmetis in all metrics but one. Moreover, \atmetis encounters OOM on Amazon.} As for the memory consumption (including the space to store hypergraphs), observe that \extendagc has comparable memory overheads with the native AHC competitors on small graphs and up to $3.1\times$ memory reduction on medium/large graphs.

\stitle{AGC} 
In Table \ref{tab:performance2} for AGC, \extendagc has comparable running time to \arw, a recent AGC method that is optimized for efficiency, while being faster than the other native AGC methods. However, the quality rank of \extendagc is 1.3, much higher than 5.6 of \arw. Specifically, in Tables \ref{tab:performance1.1} and \ref{tab:performance1.2}, \extendagc consistently achieves better clustering quality than \arw on all six attributed graphs under all metrics. 
Moreover,  \extendagc remains to be the runner-up in terms of running time on the first five datasets, and is the fastest on the largest Amazon2M for clustering. Memory-wise, \extendagc consumes a moderate amount of memory that stays below 1GB over the first four small datasets and achieves decent performance on two large datasets, TWeibo and Amazon2M.

\begin{table*}[!t]
\centering
\caption{\extension{Evaluation between \extendagc and \extendgpu.}}\label{tab:gpuexp}
\vspace{-3mm}
\extension{
\resizebox{0.92\textwidth}{!}{
\setlength{\tabcolsep}{5pt}
\renewcommand{\arraystretch}{0.93} %

\begin{tabular}{l|l|ll|ll|ll|ll|ll|ll}
\toprule
\multicolumn{1}{l|}{\multirow{2}{*}{Task}} & \multirow{2}{*}{Dataset} & \multicolumn{2}{c|}{Acc} & \multicolumn{2}{c|}{F1} & \multicolumn{2}{c|}{NMI} & \multicolumn{2}{c|}{ARI} & \multicolumn{2}{c|}{Mem} & \multicolumn{2}{c}{Time} \\ \cline{3-14} 
&  & CPU & GPU & CPU & GPU & CPU & GPU & CPU & GPU & CPU & GPU & CPU & GPU (Speedup) \\\hline
\multirow{8}{*}{AHC} & Query &0.715  &0.719  &0.662  &0.664  &0.645  &0.666  &0.571  &0.578 &0.161  &1.083 &0.342  &0.230 (1.49$\times$)  \\
 & Cora-CA &0.651  &0.653  &0.608  &0.610  &0.462  &0.469  &0.406  &0.411  &0.231  &1.096  &0.402  &0.265 (1.52$\times$) \\
& Cora-CC &0.592  &0.580  &0.520  &0.535  &0.412  &0.395  &0.338  &0.311  &0.232  &1.098  &0.416  &0.296 (1.41$\times$) \\
 & Citeseer &0.662  &0.668  &0.615  &0.620  &0.392  &0.387  &0.397  &0.410  &0.317  &1.128  &0.635  &0.575 (1.10$\times$) \\
 & 20News &0.712  &0.712  &0.658  &0.666  &0.409  &0.407  &0.469  &0.465  &0.383  &1.094  &8.176  &0.268 (30.5$\times$)  \\
 & DBLP &0.797  &0.808  &0.774  &0.787  &0.632 &0.643  &0.632  &0.646  &0.998  &1.321  &41.50  &0.591 (70.2$\times$)  \\
& Amazon &0.660  &0.648  &0.492  &0.487  &0.630  &0.636  &0.524  &0.509  &56.71  &11.16  &1286  &152.3 (8.44$\times$) \\
 & MAG-PM &0.566  &0.559  &0.405  &0.393  &0.561  &0.545  &0.471  &0.454  &59.25  &11.35  &1371  &120.2 (11.4$\times$) \\\midrule
\multirow{6}{*}{AGC} & Cora &0.723  &0.683  &0.686  &0.621  &0.556  &0.533  &0.484  &0.470  &0.369  &1.120  &1.251  &0.213 (5.87$\times$) 
\\
 & Citeseer-UG &0.691  &0.690  &0.651  &0.649  &0.438  &0.437  &0.450  &0.451  &0.517  &1.153  &1.587  &0.507 (3.13$\times$) 
 \\
 & Wiki &0.551  &0.560  &0.467  &0.487  &0.543  &0.547  &0.353  &0.368  &0.706  &1.151  &0.907  &0.357 
 (2.57$\times$) 
 \\
 & Citeseer-DG &0.696  &0.694  &0.651  &0.652  &0.444  &0.441  &0.460  &0.454  &0.280  &1.159  &0.838  &0.508 (1.65$\times$)  \\
 & TWeibo &0.433  &0.434  &0.129  &0.126  &0.023  &0.022  &0.019  &0.016  &19.89  &16.73  &1318  &105.0 (12.6$\times$)  \\
 & Amazon2M &0.494  &0.496  &0.191  &0.194  &0.441  &0.437  &0.545  &0.544  &17.01  &18.08  &1708  &158.9 (10.8$\times$) \\\midrule
\multirow{3}{*}{AMGC} & ACM &0.928  &0.924  &0.928  &0.924  &0.739  &0.730  &0.796  &0.786  &1.062  &1.267  &1.738  &0.190 (9.15$\times$) \\
 & IMDB &0.576  &0.553  &0.544 &0.510  &0.176  &0.166  &0.195  &0.184  &1.485  &1.136  &1.574  &0.236 (6.67$\times$) \\
 & DBLP-MG &0.933  &0.935  &0.929  &0.931  &0.785  &0.791  &0.839  &0.842  &0.691  &1.787  &3.766  &0.587 (6.42$\times$)\\\bottomrule
\end{tabular}}
}
\vspace{-2mm}
\end{table*}

\stitle{AMGC}
In Table \ref{tab:performance2.1}, \extendagc achieves a significant speedup ratio over the native AMGC baselines, often by an order of magnitude, while being memory efficient. Specifically, \extendagc achieves a speedup of $15.0\times$, $13.9\times$, and $9.5\times$, compared to the runner-up native AMGC methods \magc and \grace.
The memory consumption of \extendagc is also less than the majority of existing {native AMGC} methods. %

\subsubsection{Evaluation on {\normalfont \extendgpu} }\label{sec:exp:gpu}
We compare the cluster quality and efficiency of the CPU-based \extendagc against  \extendgpu in Section \ref{sec:extendgppu}, with results reported in Table \ref{tab:gpuexp} for the three tasks (AHC, AGC, and AMGC) over all datasets. 
First, observe that \extendgpu achieves similarly high-quality cluster results as the CPU-based \extendagc across all datasets for all three tasks, and the quality difference between \extendgpu and \extendagc are often negligible, in terms of Acc, F1, NMI, and ARI.

The last column of Table \ref{tab:gpuexp} provides the running time of \extendgpu and \extendagc with 16 CPU threads.
For the AHC task, the speedup of \extendgpu is less significant on the small attributed hypergraphs (Query, Cora-CA, Cora-CC, and Citeseer). We ascribe this to the numerous SVD operations on small $k\times k$ matrices in \discrgpu, as it has been known that small dimensions of input matrices may hurt the efficiency of GPU-based SVD \cite{An2016EfficientOJ}.
On medium/large attributed hypergraphs (20News, DBLP, Amazon, and MAG-PM), 
the GPU-accelerated version, \extendgpu, achieves speedup ratios of 30.5, 70.2, 8.44, and 11.4, respectively, over the CPU version \extendagc. 
The high speedup ratios of \extendgpu, often exceeding an order of magnitude, validate the efficiency of the technical designs elaborated in Section \ref{sec:extendgppu}, especially on large-scale hypergraphs.
For the AGC task, similarly, on small attributed graphs, Cora, Citeseer-UG, Wiki, and Citeseer-DG, \extendgpu is faster than \extendagc while the speedup ratio is usually below 10, due to the same reason explained above. On large attributed graphs (TWeibo and Amazon2M), \extendgpu is more efficient than \extendagc by an order of magnitude.
For the AMGC task, \extendgpu is also consistently faster than \extendagc on all attributed multiplex graphs.
The memory consumption of \extendgpu is measured by GPU video memory (VRAM), while that of \extendagc is by RAM, and the consumption is reported in the second last column of Table \ref{tab:gpuexp} in GBs. The memory usage of \extendgpu and \extendagc is not directly comparable, due to the different computational architectures and libraries used on GPUs and CPUs. 
Note that the major memory consumption of our implementations is in the KNN augmentation step. 
On small or medium-sized datasets, e.g., Query and Cora-CA, VRAM usage by \extendgpu is higher than the RAM usage by \extendagc. The reason is that \extendgpu uses GPU-based Faiss for nearest-neighbor search and Faiss allocates about 700MB of VRAM  for temporary storage.
On large datasets,   \extendagc requires a substantial RAM space due to the implementation of the ScaNN algorithm for KNN, while GPU-based Faiss in \extendgpu requires less VRAM space.

\input{gpubaseline}

\revision{
Then we enhance \grace~\citep{FanseuKamhoua2022GRACEAG} with GPU acceleration using CuPy  and cuML  libraries, resulting in \gracegpu for comparison. We also compare with the GPU-based implementation of the {S}pectral {M}odularity {M}aximization~\cite{newmanSpectralMethodsCommunity2013} clustering method dubbed as \cugraph, which operates on the graph adjacency matrix for AGC (or clique expansion of the hypergraph for AHC, or the sum of multiplex adjacency matrices for AMGC) with the attribute KNN augmentation. The results for AHC, AGC, and AMGC are presented in Tables \ref{tab:additional-gpubaselines-ahc}-\ref{tab:additional-gpubaselines-amgc}, respectively. On the first six smaller datasets in Table \ref{tab:additional-gpubaselines-ahc} for AHC, \cugraph exhibits lower quality in terms of Acc, F1, NMI, and ARI, despite comparable efficiency to \extendgpu, which delivers significantly better clustering quality. \extendgpu outperforms \gracegpu in both quality and efficiency across all AHC datasets. Notably, on large datasets Amazon and MAG-PM in Table \ref{tab:additional-gpubaselines-ahc}, \extendgpu efficiently produces satisfactory clusters, whereas \gracegpu and \cugraph encounter out-of-memory due to their requirement to expand hypergraphs into graphs. 
Similar observations are made for AGC and AMGC in Tables \ref{tab:additional-gpubaselines-agc} and \ref{tab:additional-gpubaselines-amgc}.
Similar patterns are observed for AGC and AMGC in Tables \ref{tab:additional-gpubaselines-agc} and \ref{tab:additional-gpubaselines-amgc}.
In these tasks, \extendgpu delivers superior clustering quality and efficiency on most datasets, except IMDB where \extendgpu is the second best, while \cugraph yields lower-quality outcomes and \gracegpu falls behind our method in speed. We conclude that \extendgpu offers high clustering quality with remarkable efficiency.
}

\input{figs-knnk.tex}

\subsection{Experimental Analysis}\label{sec:param-exp}

\input{figs-beta.tex}

\extension{
\input{figs-knnk-add}
}

\noindent\textbf{Varying $K$}. Figure \ref{fig:knnk} depicts the Acc, F1, NMI scores, and the {KNN computation} time of \extendagc on 8 attributed hypergraphs (AHC) when varying $K$ from $2$ to $1000$. We can make the following observations. {First, on most hypergraphs, the clustering accuracies of \extendagc first grow when $K$ is increased from $2$ to $10$ and then decline, especially when $K$ is beyond $50$.} The reasons are as follows. When $K$ is small, the KNN graph $\mathcal{G}_K$ in \extendagc fails to capture the key information in the attribute matrix $\XM$, leading to limited result quality. On the other hand, when $K$ is large, more noisy or distorted information will be introduced in $\mathcal{G}_K$, and hence, causes accuracy loss. This coincides with our observation in the preliminary study in Figure \ref{fig:sim-dist}.
{Moreover, as $K$ goes up, the time of KNN construction increases on all datasets.} %
Figures \ref{fig:knnk-add-agc} and  \ref{fig:knnk-add-amgc} show the Acc, F1, NMI scores and KNN computation time of \extendagc on the 6 attributed graphs and 3 attributed multiplex graphs for AGC and AMGC, respectively, when varying $K$ from 2 to 1000. 
On small graphs in Figure \ref{fig:knnk-add-cora-agc}-\ref{fig:knnk-add-dciteseer-agc} and Figure \ref{fig:knnk-add-amgc}, the cluster quality increases from 2 to 50, and then declines on datasets such as Citeseer-UG, Wiki, and ACM.
On large datasets TWeibo and Amazon2M in Figure \ref{fig:knnk-add-tweibo-agc} and \ref{fig:knnk-add-amazon2m-agc}, a turning point appears around $K=10$.
Therefore, we set $K$ to be 50 and 10 on these small and large datasets, respectively. 

\extension{
\input{figs-beta-add.tex}
}

\stitle{Varying $\beta$} Recall that in the generalized $(\alpha,\beta,\gamma)$-random walk model, the parameter $\beta$ is used to balance the combination of topological proximities from graph topology $\mathcal{N}_O$ and the attribute similarities from KNN graph $\mathcal{G}_K$. Figure \ref{fig:beta} displays the AHC performance of \extendagc on 8 attributed hypergraph datasets when $\beta$ varies from $0$ to $1$. When $\beta=0$, \extendagc degrades to a hypergraph clustering method without the consideration of any attribute information, whereas \extendagc only clusters the KNN graph $\mathcal{G}_K$ regardless of the topology structure in $\mathcal{H}$ if $\beta=1$. {From Figure \ref{fig:beta}, we can see a large $\beta$ (e.g., $0.7$-$0.8$) on small/medium datasets (Query, Cora-CA, Cora-CC, Citeseer, 20News, and DBLP) bring more performance enhancements, meaning that attribute information plays  more important roles on those datasets. 
This is because they have limited amounts of connections (or are too dense to be informative, e.g., on Query) in the original hypergraph structure as listed in Table \ref{tab:dataset} and rely on attribute similarities from the augmented KNN graph $\mathcal{G}_K$ for improved clustering. By contrast, on Amazon and MAG-PM, \extendagc achieves the best clustering quality with small $\beta$ in $[0.1,0.4]$, indicating graph topology has higher weights on Amazon and MAG-PM.}
\extension{Figures \ref{fig:beta-add-agc} and \ref{fig:beta-add-amgc} report the Acc, F1, and NMI scores on AGC and AMGC tasks respectively. Similarly, when $\beta$ increases from 0, the cluster quality generally improves, then becomes stable around 0.4 and 0.5, and decreases when $\beta$ is large and close to $1$. 
On DBLP-MG in Figure \ref{fig:beta-add-amgc}, the highest clustering quality can be acquired with a small $\beta$ around 0.1. We infer that node attributes in this dataset are of limited significance for clustering, while on ACM and IMDB, the best quality is achieved when $\beta$ appropriately balances graph topology and attributes.}

\input{figs-gamma.tex}

\input{random-init.tex}

\input{random-init-extension}

\stitle{Varying $\gamma$} We evaluate \extendagc in terms of AHC quality and running time when varying $\gamma$. Figure \ref{fig:gamma} displays the Acc, F1, NMI, and time on two representative datasets when $\gamma$ varies from 1 to 5. The results on other datasets are similar and thus are omitted for space. Observe that in practice the Acc, F1, and NMI scores obtained by \extendagc first increase and then remain stable when $\gamma$ is beyond 3 and 2 on Cora-CC and Citeseer, respectively.
By contrast, the running time goes up as $\gamma$ increases. Therefore,  we set $\gamma=3$ in experiments.

\vspace{-1mm}
\stitle{Effectiveness Evaluation of InitBCM and Discretize} 
On attributed hypergraphs, to verify the effectiveness of  $\bcm$ for the BCM initialization, we compare \extendagc with the ablated version \extendagc-random-init, where the BCM matrix $\YM^{(0)}$ is initialized at random. In Table \ref{tab:ablation}, \extendagc obtains remarkable improvements over \extendagc-random-init in Acc, F1, and NMI in comparable processing time. For instance, on Amazon, \extendagc outperforms \extendagc-random-init by a large margin of 3.7\% Acc, 19.5\% F1, and 6.8\% NMI with 24 seconds less to process. {On MAG-PM, \extendagc needs additional time compared to \extendagc-random-init. The reason is that \extendagc-random-init 
starts with a low-quality BCM and converges to local optimum solutions with suboptimal MHC,
whereas \extendagc can bypass such pitfalls with a good initial BCM from \bcm{} and continue searching for the optimal solution with more iterations,
which in turn results in a considerable gap in clustering quality.}
In addition, we validate the effectiveness of \discr{} used in \extendagc to transform $k$ leading eigenvectors $\QM$ to BCM matrix $\YM$. Table \ref{tab:ablation} reports the  accuracy of  \extendagc and a variant \extendagc-\kmeans obtained by replacing \discr in \extendagc with \kmeans on all datasets. It can be observed that compared with \extendagc-\kmeans, \extendagc is able to output high-quality BCM matrices $\YM$ with substantially higher clustering accuracy scores while being up to $3.2\times$ faster. 
\revision{
The ablation results on AGC and AMGC are in Tables \ref{tab:ablation_extension_agc} and \ref{tab:ablation_extension_amgc}, respectively. Regarding clustering quality (Acc, F1, NMI), Table \ref{tab:ablation_extension_agc} shows that for AGC, \extendagc surpasses its ablated counterparts on all datasets across most effectiveness metrics, except for the Citeseer datasets. For example, \extendagc with \bcm achieves an Acc that is 4.2\% higher than \extendagc-random-init on Amazon2M. In  Table \ref{tab:ablation_extension_amgc} for AMGC, \extendagc performs the best on all the three datasets. For efficiency  in Tables \ref{tab:ablation_extension_agc} and \ref{tab:ablation_extension_amgc}, \extendagc is similar to \extendagc-random-init, while \extendagc-\kmeans is slower. These results confirm the effectiveness of the proposed techniques for AGC and AMGC.
}

\input{figs-convergence.tex}

\extension{
\input{fig-runtime}
}

\subsection{Convergence Analysis}\label{sec:cvg-als}
We provide an empirical analysis pertinent to the convergence of \extendagc for attributed hypergraph clustering. To do so, we first disable the early termination strategies at Line 10 in Algorithm \ref{alg:mainalg}. We also set $\tau=1$ so as to evaluate the MHC (denoted as $\phi_t$) of the BCM matrix $\YM^{(t)}$ generated in each $t$-th iteration of \extendagc and \extendagc-random-init, where $t$ starts from 0 till convergence. Furthermore, we calculate the Acc, F1, and NMI scores with the ground truth for each BCM matrix $\YM^{(t)}$ generated throughout the iterative procedures of \extendagc. 
Figure \ref{fig:convergence} shows the MHC $\phi_t$, Acc, F1, and NMI scores based on the BCM matrix of each iteration in \extendagc, as well as the MHC of \extendagc-random-init over all datasets. Notably, {MHC $\phi_t$} experiences a sharp decline when $t$ increases from 0 to 50 on most hypergraphs, while the Acc, F1, and NMI results have significant growth. Moreover, compared to MHC with random init, MHC curves of \extendagc are mostly lower (better) on all datasets under the same $t$-th iteration. These phenomena demonstrate the effectiveness of \bcm{} in facilitating fast convergence of \extendagc. 
However, when we keep increasing $t$, these scores either remain stable or deteriorate. For instance, MHC scores grow significantly after $10$ iterations on Amazon, while there is a big drop in Acc and F1 scores when $t\ge 45$ on DBLP. This indicates that adding more iterations does not necessarily ensure better solutions. Hence, the early termination proposed in \extendagc can serve as an effective approach to remedy this issue.

\subsection{Runtime Analysis} \label{sec:runtime-als}
Figure \ref{fig:runtime-vs} reports time breakdown of \extendagc and \extendgpu into four parts:  KNN construction, orthogonal iterations, discretization, and greedy initialization and MHC evaluation on all attributed hypergraphs. 
We first explain the results of \extendagc on CPUs. On all datasets, the four parts in \extendagc all take considerable time to process, except 20News and DBLP, where KNN construction dominates, since 20News and DBLP contain many nodes but relatively few edges.
\extension{Then, we compare the time breakdown of \extendgpu with \extendagc. 
On small attributed hypergraphs (Query, Cora-CA, Cora-CC, and Citeseer) in Figures \ref{fig:runtime-query}, \ref{fig:runtime-cora-coauth}, \ref{fig:runtime-cora-cocite}, and \ref{fig:runtime-citeseer-cocite}, observe that \extendgpu significantly reduces the time for KNN, while the other time costs are on par with that of \extendagc, which is consistent with the results in Section \ref{sec:exp:gpu}. 
On medium-sized/large attributed hypergraphs in Figures \ref{fig:runtime-20news}, \ref{fig:runtime-dblp-coauth}, \ref{fig:runtime-amazon}, and \ref{fig:runtime-mag}, \extendgpu significantly improves the efficiency on all of KNN construction, orthogonal iterations, discretization,  greedy initialization and MHC evaluation. 
\revision{From the results on Amazon and MAG-PM, we observe that the  scalability of \extendgpu is primarily constrained by KNN construction, while the overhead of the CPU-based \extendagc is more evenly distributed across the four parts.}
}

%% file: figs-knnk.tex
\begin{figure*}[!t]
\centering
\definecolor{darkgray176}{RGB}{176,176,176}
\definecolor{darkorange25512714}{RGB}{255,127,14}
\definecolor{darkviolet1910191}{RGB}{191,0,191}
\definecolor{forestgreen4416044}{RGB}{44,160,44}
\definecolor{lightgray204}{RGB}{204,204,204}
\definecolor{steelblue31119180}{RGB}{31,119,180}
\definecolor{tabred}{RGB}{214, 39, 40}

\begin{tikzpicture}
    \begin{customlegend}[
        legend entries={Acc, F1, NMI, time},
        legend columns=4,
        legend style={at={(0.5,1.05)},anchor=north,draw=none,font=\small,column sep=0.2cm}]
        \addlegendimage{line width=2pt, color=steelblue31119180, dashed}
        \addlegendimage{line width=2pt, color=darkorange25512714, dotted}
        \addlegendimage{line width=2pt, color=forestgreen4416044}
        \addlegendimage{line width=2pt, color=darkviolet1910191}
    \end{customlegend}
\end{tikzpicture}
\vspace{-2mm}
\\[-\lineskip]

\captionsetup[subfloat]{labelfont={Large}, textfont={Large}}
\captionsetup{justification=centering}
\resizebox{0.22\textwidth}{!}{%
\subfloat[{Query}]{
\label{fig:knnk-query}
\input{knnk-query.tex}
}}
\resizebox{0.22\textwidth}{!}{%
\subfloat[Cora-CA]{
\label{fig:knnk-cora-ca}
\input{knnk-cora-coauth.tex}
}}
\resizebox{0.22\textwidth}{!}{%
\subfloat[Cora-CC]{
\label{fig:knnk-cora-cc}
\input{knnk-cora-cocite.tex}
}}
\resizebox{0.22\textwidth}{!}{%
\subfloat[Citeseer]{
\label{fig:knnk-citeseer-cc}
\input{knnk-citeseer-cocite.tex}
}}\vspace{-2mm}

\resizebox{0.22\textwidth}{!}{%
\subfloat[{20News}]{
\label{fig:knnk-20news}
\input{knnk-20news.tex}
}}
\resizebox{0.22\textwidth}{!}{%
\subfloat[DBLP]{
\label{fig:knnk-dblp-ca}
\input{knnk-dblp-coauth.tex}
}}
\resizebox{0.22\textwidth}{!}{%
\subfloat[Amazon]{
\label{fig:knnk-amazon}
\input{knnk-amazon.tex}
}}
\resizebox{0.22\textwidth}{!}{%
\subfloat[MAG-PM]{
\label{fig:knnk-mag}
\input{knnk-mag.tex}
}}\vspace{-3mm}
\caption{{Varying $K$ for AHC (best viewed in color).}} \label{fig:knnk}
\vspace{-3mm}
\end{figure*}

%% file: figs-beta.tex
\begin{figure*}[!t]
\centering
\definecolor{darkgray176}{RGB}{176,176,176}
\definecolor{darkorange25512714}{RGB}{255,127,14}
\definecolor{forestgreen4416044}{RGB}{44,160,44}
\definecolor{lightgray204}{RGB}{204,204,204}
\definecolor{steelblue31119180}{RGB}{31,119,180}

\begin{tikzpicture}
    \begin{customlegend}[
        legend entries={Acc, F1, NMI},
        legend columns=3,
        legend style={at={(0.5,1.05)},anchor=north,draw=none,font=\small,column sep=0.2cm}]
        \addlegendimage{line width=2pt, color=steelblue31119180, dashed}
        \addlegendimage{line width=2pt, color=darkorange25512714, dotted}
        \addlegendimage{line width=2pt, color=forestgreen4416044}
    \end{customlegend}
\end{tikzpicture}
\vspace{-2mm}
\\[-\lineskip]
\captionsetup[subfloat]{labelfont={huge}, textfont={huge}}
\showcaptionsetup{subfloat}

\resizebox{0.121\textwidth}{!}{
\subfloat[Query]{
\label{fig:beta-query}
\input{beta-query.tex}
}}
\resizebox{0.121\textwidth}{!}{%
\subfloat[Cora-CA]{
\label{fig:beta-cora-ca}
\input{beta-cora-ca.tex}
}}
\resizebox{0.121\textwidth}{!}{%
\subfloat[Cora-CC]{
    \label{fig:beta-cora-cc}
    \input{beta-cora-cc.tex}
}}
\resizebox{0.121\textwidth}{!}{%
\subfloat[Citeseer]{
    \label{fig:beta-citeseer-cc}
    \input{beta-citeseer.tex}
}}
\resizebox{0.121\textwidth}{!}{%
\subfloat[{20News}]{
\label{fig:beta-20news}
\input{beta-20news.tex}
}}
\resizebox{0.121\textwidth}{!}{%
\subfloat[DBLP]{
    \label{fig:beta-dblp-ca}
    \input{beta-dblp.tex}
}}
\resizebox{0.121\textwidth}{!}{%
\subfloat[Amazon]{
    \label{fig:beta-amazon}
    \input{beta-amazon.tex}
}}
\resizebox{0.121\textwidth}{!}{%
\subfloat[MAG-PM]{
    \label{fig:beta-mag}
    \input{beta-mag.tex}
}}
\vspace{-2mm}
\caption{{Varying $\beta$ for AHC (best viewed in color).}} \label{fig:beta}
\vspace{-2mm}

\end{figure*}

%% file: figs-knnk-add.tex
\begin{figure}[!h]
\centering
\definecolor{darkgray176}{RGB}{176,176,176}
\definecolor{darkorange25512714}{RGB}{255,127,14}
\definecolor{darkviolet1910191}{RGB}{191,0,191}
\definecolor{forestgreen4416044}{RGB}{44,160,44}
\definecolor{lightgray204}{RGB}{204,204,204}
\definecolor{steelblue31119180}{RGB}{31,119,180}
\definecolor{tabred}{RGB}{214, 39, 40}

\begin{tikzpicture}
    \begin{customlegend}[
        legend entries={Acc, F1, NMI, time},
        legend columns=4,
        legend style={at={(0.5,1.05)},anchor=north,draw=none,font=\small,column sep=0.2cm}]
        \addlegendimage{line width=2pt, color=steelblue31119180, dashed}
        \addlegendimage{line width=2pt, color=darkorange25512714, dotted}
        \addlegendimage{line width=2pt, color=forestgreen4416044}
        \addlegendimage{line width=2pt, color=darkviolet1910191}
    \end{customlegend}
\end{tikzpicture}
\vspace{-2mm}
\\[-\lineskip]

\captionsetup[subfloat]{labelfont={LARGE}, textfont={LARGE}}
\captionsetup{justification=centering}
\resizebox{0.155\textwidth}{!}{%
\subfloat[{Cora}]{
\label{fig:knnk-add-cora-agc}
\input{knnk-add-cora}
}}
\resizebox{0.155\textwidth}{!}{%
\subfloat[Citeseer-UG]{
\label{fig:knnk-add-citeseer}
\input{knnk-add-citeseer}
}}
\resizebox{0.155\textwidth}{!}{%
\subfloat[Wiki]{
\label{fig:knnk-add-wiki}
\input{knnk-add-wiki}
}}
\vspace{-1mm}

\hspace{-1mm}
\resizebox{0.155\textwidth}{!}{%
\subfloat[Citeseer-DG]{
\label{fig:knnk-add-dciteseer-agc}
\input{knnk-add-dciteseer}
}}
\resizebox{0.16\textwidth}{!}{%
\subfloat[Tweibo]{
\label{fig:knnk-add-tweibo-agc}
\input{knnk-add-tweibo}
}}
\resizebox{0.16\textwidth}{!}{%
\subfloat[Amazon2M]{
\label{fig:knnk-add-amazon2m-agc}
\input{knnk-add-amazon2m}
}}
\vspace{-3mm}
\caption{\extension{Varying $K$ for AGC (best viewed in color).}} \label{fig:knnk-add-agc}
\vspace{-3mm}
\end{figure}

\begin{figure}[!h]
\centering
\definecolor{darkgray176}{RGB}{176,176,176}
\definecolor{darkorange25512714}{RGB}{255,127,14}
\definecolor{darkviolet1910191}{RGB}{191,0,191}
\definecolor{forestgreen4416044}{RGB}{44,160,44}
\definecolor{lightgray204}{RGB}{204,204,204}
\definecolor{steelblue31119180}{RGB}{31,119,180}
\definecolor{tabred}{RGB}{214, 39, 40}

\captionsetup[subfloat]{labelfont={LARGE}, textfont={LARGE}}
\vspace{-3mm}
\hspace{-1.5mm}
\resizebox{0.155\textwidth}{!}{%
\subfloat[ACM]{
\label{fig:knnk-add-acm}
\input{knnk-add-acm}
}}
\hspace{-1mm}
\resizebox{0.155\textwidth}{!}{%
\subfloat[{IMDB}]{
\label{fig:knnk-add-imdb}
\input{knnk-add-imdb}
}}
\hspace{-1mm}
\resizebox{0.155\textwidth}{!}{%
\subfloat[DBLP-MG]{
\label{fig:knnk-add-dblp}
\input{knnk-add-dblp}
}}
\vspace{-3mm}
\caption{\extension{Varying $K$ for AMGC (best viewed in color).}} \label{fig:knnk-add-amgc}
\vspace{-3mm}
\end{figure}

%% file: figs-beta-add.tex
\begin{figure}[!t]
\centering
\definecolor{darkgray176}{RGB}{176,176,176}
\definecolor{darkorange25512714}{RGB}{255,127,14}
\definecolor{forestgreen4416044}{RGB}{44,160,44}
\definecolor{lightgray204}{RGB}{204,204,204}
\definecolor{steelblue31119180}{RGB}{31,119,180}

\begin{tikzpicture}
    \begin{customlegend}[
        legend entries={Acc, F1, NMI},
        legend columns=3,
        legend style={at={(0.5,1.05)},anchor=north,draw=none,font=\small,column sep=0.2cm}]
        \addlegendimage{line width=2pt, color=steelblue31119180, dashed}
        \addlegendimage{line width=2pt, color=darkorange25512714, dotted}
        \addlegendimage{line width=2pt, color=forestgreen4416044}
    \end{customlegend}
\end{tikzpicture}
\vspace{-2mm}
\\[-\lineskip]
\captionsetup[subfloat]{labelfont={LARGE}, textfont={LARGE}}
\showcaptionsetup{subfloat}

\resizebox{0.145\textwidth}{!}{
\subfloat[Cora]{
\label{fig:beta-add-cora}
\input{beta-add-cora}
}}
\resizebox{0.145\textwidth}{!}{%
\subfloat[Citeseer-UG]{
\label{fig:beta-add-citeseer}
\input{beta-add-citeseer}
}}
\resizebox{0.145\textwidth}{!}{%
\subfloat[Wiki]{
    \label{fig:beta-add-wiki}
    \input{beta-add-wiki}
}}
\vspace{-2mm}

\resizebox{0.145\textwidth}{!}{%
\subfloat[Citeseer-DG]{
    \label{fig:beta-add-dciteseer}
    \input{beta-add-dciteseer}
}}
\resizebox{0.145\textwidth}{!}{%
\subfloat[Tweibo]{
    \label{fig:beta-add-tweibo}
    \input{beta-add-tweibo}
}}
\resizebox{0.145\textwidth}{!}{%
\subfloat[Amazon2M]{
    \label{fig:beta-add-amazon2m}
    \input{beta-add-amazon2m}
}}
\vspace{-3mm}
\caption{\extension{Varying $\beta$ for AGC (best viewed in color).}} \label{fig:beta-add-agc}
\vspace{-5mm}
\end{figure}

\begin{figure}[!t]
\centering
\definecolor{darkgray176}{RGB}{176,176,176}
\definecolor{darkorange25512714}{RGB}{255,127,14}
\definecolor{forestgreen4416044}{RGB}{44,160,44}
\definecolor{lightgray204}{RGB}{204,204,204}
\definecolor{steelblue31119180}{RGB}{31,119,180}

\captionsetup[subfloat]{labelfont={LARGE}, textfont={LARGE}}
\showcaptionsetup{subfloat}
\resizebox{0.145\textwidth}{!}{%
\subfloat[ACM]{
    \label{fig:beta-add-acm}
    \input{beta-add-acm}
}}
\resizebox{0.14\textwidth}{!}{%
\subfloat[IMDB]{
    \label{figs/beta-add-imdb}
    \input{beta-add-imdb}
}}
\resizebox{0.14\textwidth}{!}{%
\subfloat[DBLP-MG]{
    \label{fig:beta-add-dblp}
    \input{beta-add-dblp}
}}

\vspace{-2mm}

\caption{{\extension{Varying $\beta$ for AMGC (best viewed in color).}}
} \label{fig:beta-add-amgc}
\vspace{-2mm}

\end{figure}

%% file: figs-gamma.tex
\begin{figure}[]
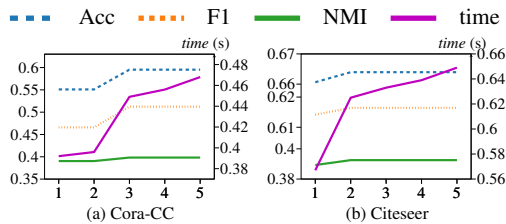

\centering
\begin{small}
\definecolor{darkgray176}{RGB}{176,176,176}
\definecolor{darkorange25512714}{RGB}{255,127,14}
\definecolor{darkviolet1910191}{RGB}{191,0,191}
\definecolor{forestgreen4416044}{RGB}{44,160,44}
\definecolor{lightgray204}{RGB}{204,204,204}
\definecolor{steelblue31119180}{RGB}{31,119,180}
\definecolor{tabred}{RGB}{214, 39, 40}

\begin{tikzpicture}
    \begin{customlegend}[
        legend entries={Acc, F1, NMI, time},
        legend columns=4,
        legend style={at={(0.5,1.05)},anchor=north,draw=none,font=\small,column sep=0.2cm}]
        \addlegendimage{line width=2pt, color=steelblue31119180, dashed}
        \addlegendimage{line width=2pt, color=darkorange25512714, dotted}
        \addlegendimage{line width=2pt, color=forestgreen4416044}
        \addlegendimage{line width=2pt, color=darkviolet1910191}
    \end{customlegend}
\end{tikzpicture}
\vspace{-2mm}
\\[-\lineskip]

\captionsetup[subfloat]{labelfont={LARGE}, textfont={LARGE}}
\resizebox{0.19\textwidth}{!}{%
\subfloat[Cora-CC]{
\label{fig:gamma-cora-cc}
\input{gamma-cora-cc.tex}
}}
\resizebox{0.19\textwidth}{!}{%
\subfloat[Citeseer]{
\label{fig:gamma-citeseer}
\input{gamma-citeseer-scale.tex}
}}

\end{small}
\vspace{-3mm}
\caption{{Varying $\gamma$ on Attributed Hypergraphs.}} \label{fig:gamma}
\vspace{-3mm}
\end{figure}

%% file: figs-convergence.tex
\begin{figure*}[!t]
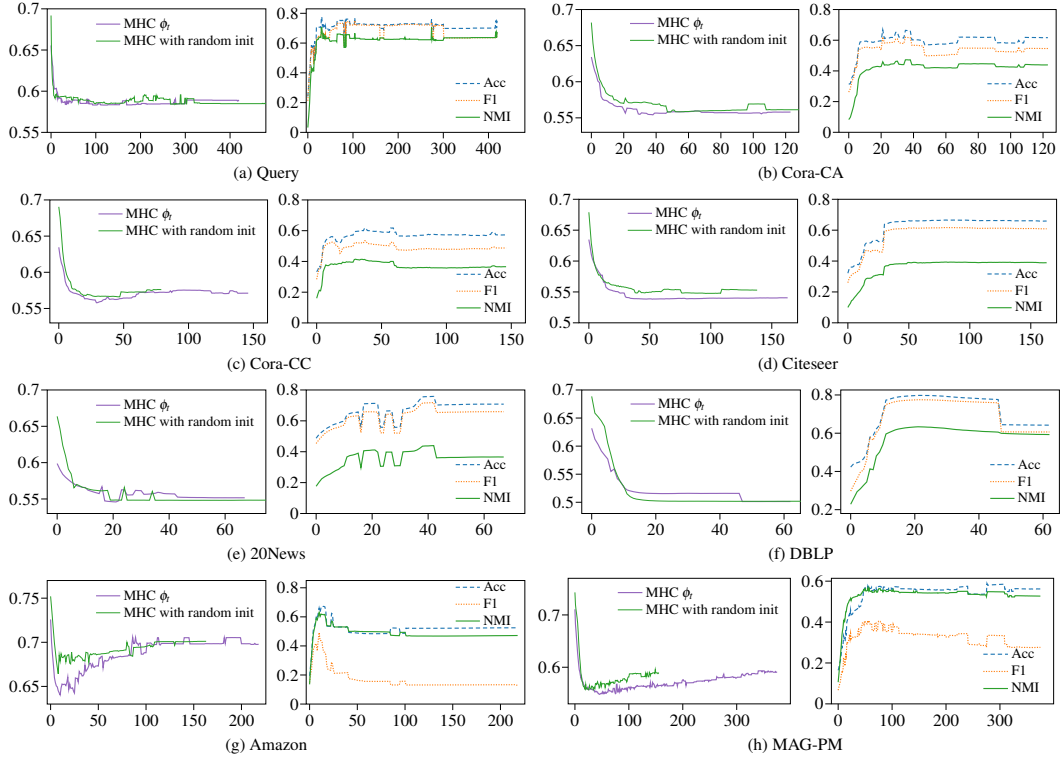

\centering
\begin{small}
\captionsetup[subfloat]{labelfont={large}, textfont={large}}
\resizebox{0.4\textwidth}{!}{%
\subfloat[{Query}]{
    \label{conv-query}
\input{convergence-query.tex}
}}
\resizebox{0.4\textwidth}{!}{%
\subfloat[Cora-CA]{
    \label{fig:conv-cora-ca}
    \input{convergence-cora-ca.tex}
}}%
\vspace{-1.5mm}

\resizebox{0.4\textwidth}{!}{%
\subfloat[Cora-CC]{
    \label{conv-cora-cc}
\input{convergence-cora-cc.tex}
}}%
\resizebox{0.4\textwidth}{!}{%
\subfloat[Citeseer]{
    \label{fig:conv-citeseer}
    \input{convergence-citeseer.tex}
}}%
\vspace{-1.5mm}

\resizebox{0.4\textwidth}{!}{%
\subfloat[{20News}]{
    \label{conv-20news}
\input{convergence-20news.tex}
}}
\resizebox{0.4\textwidth}{!}{%
\subfloat[DBLP]{
    \label{fig:conv-dblp-ca}
    \input{convergence-dblp-ca.tex}
}}%
\vspace{-1.5mm}

\resizebox{0.4\textwidth}{!}{%
\subfloat[Amazon]{
    \label{fig:conv-amazon}
    \input{convergence-amazon.tex}
}}%
\resizebox{0.4\textwidth}{!}{%
\subfloat[MAG-PM]{
    \label{fig:conv-mag}
    \input{convergence-mag.tex}
}}%
\end{small}
\vspace{-3mm}
\caption{{Convergence Analysis (best viewed in color).} 
} \label{fig:convergence}
\vspace{-2mm}
\end{figure*}

%% file: fig-runtime.tex
\begin{figure*}[t]
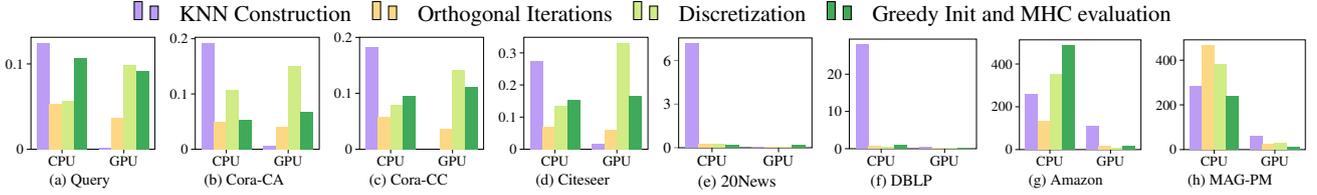

\centering
\definecolor{darkgray176}{RGB}{176,176,176}
\definecolor{khaki209235133}{RGB}{209,235,133}
\definecolor{khaki253216132}{RGB}{253,216,132}
\definecolor{mediumseagreen6317089}{RGB}{63,170,89}
\definecolor{tomato2297752}{RGB}{188, 157, 245}

\begin{tikzpicture}
    \begin{customlegend}[
        legend entries={KNN Construction, Orthogonal Iterations, Discretization, Greedy Init and MHC evaluation},
        legend columns=4,
        legend style={at={(0.5,1.05)},anchor=north,draw=none,font=\small,column sep=0.2cm}]
        \addlegendimage{ybar,ybar legend,draw=none,fill=tomato2297752}
        \addlegendimage{ybar,ybar legend,draw=none,fill=khaki253216132}
        \addlegendimage{ybar,ybar legend,draw=none,fill=khaki209235133}
        \addlegendimage{ybar,ybar legend,draw=none,fill=mediumseagreen6317089}
    \end{customlegend}
\end{tikzpicture}
\vspace{-2mm}
\\[-\lineskip]

\captionsetup[subfloat]{labelfont={LARGE}, textfont={LARGE}}
\captionsetup{justification=centering}
\resizebox{0.12\textwidth}{!}{%
\subfloat[{Query}]{
\label{fig:runtime-query}
\input{runtime-query}
}}
\captionsetup{justification=centering}
\resizebox{0.12\textwidth}{!}{%
\subfloat[{Cora-CA}]{
\label{fig:runtime-cora-coauth}
\input{runtime-cora-coauth}
}}
\captionsetup{justification=centering}
\resizebox{0.12\textwidth}{!}{%
\subfloat[{Cora-CC}]{
\label{fig:runtime-cora-cocite}
\input{runtime-cora-cocite}
}}
\captionsetup{justification=centering}
\resizebox{0.12\textwidth}{!}{%
\subfloat[{Citeseer}]{
\label{fig:runtime-citeseer-cocite}
\input{runtime-citeseer-cocite}
}}\vspace{-2mm}
\resizebox{0.12\textwidth}{!}{%
\subfloat[{20News}]{
\label{fig:runtime-20news}
\input{runtime-20news}
}}
\captionsetup{justification=centering}
\resizebox{0.12\textwidth}{!}{%
\subfloat[{DBLP}]{
\label{fig:runtime-dblp-coauth}
\input{runtime-dblp-coauth}
}}
\captionsetup{justification=centering}
\resizebox{0.12\textwidth}{!}{%
\subfloat[{Amazon}]{
\label{fig:runtime-amazon}
\input{runtime-amazon}
}}
\captionsetup{justification=centering}
\resizebox{0.12\textwidth}{!}{%
\subfloat[{MAG-PM}]{
\label{fig:runtime-mag}
\input{runtime-mag}
}}
\vspace{-1mm}
\caption{\extension{Runtime breakdown of CPU-based \extendagc and \extendgpu in seconds.}} \label{fig:runtime-vs}
\vspace{-2mm}
\end{figure*}

%% file: related_work.tex
\vspace{-1mm}
\section{Related work} \label{sec:relatedwork}
\vspace{-1mm}
\noindent{\bf Hypergraph Clustering.}
Motivated by the applications in circuit manufacturing, partitioning algorithms have been developed to divide hypergraphs into partitions/clusters, such as \texttt{hMetis}~\cite{karypisMultilevelHypergraphPartitioning1999a} and \kahypar~\cite{kahypar10.1145/3529090}. These methods typically adopt a three-stage framework consisting of coarsening, initial clustering, and refinement stages.
These algorithms directly perform clustering on a coarsened hypergraph with a relatively small size. In addition, they run a portfolio of clustering algorithms and select the best outcome. These algorithms rely on a set of clustering heuristics and lack the extensibility for exploiting node attribute information.
Hypergraph Normalized Cut (HNCut)~\cite{zhouLearningHypergraphsClustering2007} is a conductance measure for hypergraph clusters from which the normalized hypergraph Laplacian $\Delta=\IM - \Theta$ is derived for spectral clustering, where  
$
    \Theta = \DM_V^{-1/2} \HM^T \DM_E^{-1} \HM \DM_V^{-1/2}.
$
Alternatively, \texttt{hGraclus}~\cite{whangMEGAMultiviewSemisupervised2020} optimizes the HNCut objective using a multi-level kernel K-means algorithm. Non-negative matrix factorization has also been applied to hypergraph clustering~\cite{hayashi_hypergraph_2020}. Despite the theoretical soundness, these algorithms are less efficient than the aforementioned partitioning algorithms and they do not utilize node attributes either.
For the problem of hypergraph local clustering, which is to find a high-quality cluster containing a specified node, a sweep cut method is proposed~\cite{takai_hypergraph_2020} to find the cluster based on hypergraph Personalized PageRank (PPR) values.
In this paper, we focus on global clustering,  a different problem from local clustering.

\stitle{Attributed Hypergraph Clustering} There exist studies designing dedicated clustering algorithms on attributed hypergraphs.
\jnmf~\cite{duHybridClusteringBased2019} is an AHC algorithm based on non-negative matrix factorization (NMF). With normalized hypergraph Laplacian~\cite{zhouLearningHypergraphsClustering2007} matrix $\Delta=\IM - \Theta$ and attribute matrix $\XM$, \jnmf optimizes the following joint objective that includes a basic NMF part as well as a symmetric NMF part:
    $\min_{\WM, \MM, \mathbf{\tilde{M}}\geq 0} ||\XM-\WM\MM||_F^2 + \alpha ||\Theta-\mathbf{\tilde{M}}\transpose \MM||_F^2 + \beta ||\mathbf{\tilde{M}}-\MM||_F^2$. 
With optimization using block coordinate descent (BCD) scheme, the matrix $\MM$ is expected to encode cluster memberships. \texttt{MEGA}~\cite{whangMEGAMultiviewSemisupervised2020} extends the formulation of \jnmf clustering objective for {\em semi-supervised} clustering of multi-view data containing hypergraph, node attributes as well as pair-wise similarity graph. \texttt{MEGA}'s clustering performance is further enhanced by initialization with \texttt{hGraclus} algorithm.
\revision{\texttt{GNMF}~\cite{cai2010graph} algorithm is originally proposed for high dimensional data clustering, while the authors of~\cite{fanseukamhouaHyperGraphConvolutionBased2021} extend its objective with the hypergraph normalized Laplacian~\cite{zhouLearningHypergraphsClustering2007} so that it spawns baseline methods for AHC.}
Although NMF-based algorithms sometimes produce clusters with good quality, their scalability is underwhelming as shown in our experiments. 
As the state-of-the-art algorithm for attributed hypergraph clustering, \grac~\cite{fanseukamhouaHyperGraphConvolutionBased2021} performs hypergraph convolution ~\cite{yadati2019hypergcn} on node attributes, which resembles the hypergraph diffusion process with mediators~\cite{chan_generalizing_2018}. Then clusters are predicted from the propagated features via a spectral algorithm.%

\stitle{Attributed Graph Clustering}
There exists a collection of studies on attributed graph clustering. 
Some studies perform attributed graph clustering by adopting probabilistic models to combine graph structure with attributes, including discriminative models such as \texttt{PCL-DC}~\cite{yangCombiningLinkContent2009} and generative models such as \texttt{BAGC}~\cite{xuModelbasedApproachAttributed2012}. Nevertheless, these methods are typically limited to handling categorical attributes. Moreover, inference over the probability distribution of $O(2^n)$ hyperedges poses a significant challenge against their generalization to hypergraph. \extension{\texttt{GNMF}~\cite{cai2010graph} is an NMF-based algorithm that enhances performance by modifying the Laplacian regularizer used in traditional NMF to utilize the Laplacian matrix constructed from the graph structure.}
Within the random walk framework, \texttt{SA-Cluster}~\cite{zhouClusteringLargeAttributed2010} algorithm augments the original graph with virtual nodes representing each possible attribute-value pair and performs k-Medroids clustering using a random walk distance measure. \arw~\cite{yangEffectiveScalableClustering2021} defines attributed random walk by adding virtual attribute nodes as bridges and combines it with graph random walk into a joint transition matrix. 
In a fashion similar to GCN~\cite{welling2016semi}, \agc~\cite{zhangAGC10.5555/3367471.3367643} performs graph convolution on node attributes to produce smooth feature representations that incorporate network structure information and subsequently applies spectral clustering. For their spectral algorithm, the authors also design heuristics to prevent propagated features from over-smoothing that undermines cluster quality.
\extension{\grace~\cite{FanseuKamhoua2022GRACEAG} adopts graph convolution on node attributes to fuse all available information and perform a spectral algorithm based on \grac~\cite{fanseukamhouaHyperGraphConvolutionBased2021}. \fgc~\cite{Kang2022FinegrainedAG} exploits both node features
and structure information via graph convolution and applies spectral clustering on a fine-grained graph that encodes higher-order relations.}

\extension{\stitle{Attributed Multiplex Graph Clustering} 
Via unsupervised learning on attributed multiplex graphs, neural network models can learn node embeddings for clustering, e.g.,  \omac \cite{Fan2020One2MultiGA} and \hdmi \cite{Jing2021HDMIHD}. 
\grace~\cite{FanseuKamhoua2022GRACEAG} constructs a multiplex graph Laplacian and uses this matrix for graph convolution. 
Other methods find a single graph that encodes the node proximity relations in all graph layers and attributes. \texttt{MCGC}~\cite{Pan2021MultiviewCG} performs graph filtering on attributes and learns a consensus graph leveraging contrastive regularization, while \magc \cite{Lin2021MultiViewAG} exploits higher-order proximity to learn consensus graphs without deep neural networks.
}

%% file: conclusion.tex
\section{Conclusion}\label{sec:conclusion}
\extension{This paper presents \extendagc, a versatile, effective, and efficient attributed network clustering method for AHC, AGC, and AMGC computation.} The {improvements}  of \extendagc over existing solutions in terms of efficiency and effectiveness is attributed to: (i)
an effective KNN augmentation strategy to exploit useful attribute information, (ii) a novel problem formulation based on a random walk model, and (iii) an efficient iterative optimization framework with speedup techniques. 
\extension{To further boost the efficiency, we  leverage GPUs and develop \extendgpu that is faster than its CPU-parallel counterpart \extendagc on large datasets, while retaining high cluster quality.
We conduct extensive experiments over real-world data to validate the {outstanding} performance of our methods.}
\revision{In the future, we plan to extend \extendagc to cope with evolving attributed networks and enhance its scalability via distributed KNN construction and matrix computation.}